\documentclass[twocolumn,trackchanges]{aastex62}

\usepackage{amsmath}
\allowdisplaybreaks

\tabletypesize{\scriptsize}

\newcommand{\col}[1]{\textcolor{black}{#1}} 
\newcommand{\cols}[1]{\textcolor{black}{#1}} 
\newcommand{\colIK}[1]{\textcolor{black}{#1}}
\newcommand{\colFo}[1]{\textcolor{black}{#1}} 

\graphicspath{{figs/}}

\shorttitle{On the emergence of a new instability during core-collapse of very massive stars}
\shortauthors{Kalashnikov et al.}

\begin{document}

\title{On the emergence of a new instability during core-collapse of very massive stars}

\correspondingauthor{Ilia Kalashnikov} \email{kalasxel@gmail.com}

\author{Ilia Kalashnikov}
\affiliation{Keldysh Institute of Applied Mathematics, 
4 Miusskaya sq., Moscow, 125047, Russia}

\author{Andrey Baranov}
\affiliation{Institute for Computer Aided Design, Russian Academy of Sciences, 19/18 2-ya Brestskaya st., Moscow, 123056, Russia}

\author{Pascal Chardonnet}
\affiliation{SCAC/IFA Embassy of France in Algeria,
25 Chemin Abdelkader Gadouche, Hydra, 16035, Algeria\footnote{On leave of absence from: Univ. Grenoble Alpes, 
\\ USMB, CNRS, LAPTh, F-74000 Annecy}
}

\author{Valery Chechetkin}  
\affiliation{Keldysh Institute of Applied Mathematics, 
4 Miusskaya sq., Moscow, 125047, Russia}
\affiliation{Institute for Computer Aided Design, Russian Academy of Sciences, 19/18 2-ya Brestskaya st., Moscow, 123056, Russia}

\author{Anastasia Filina}
\affiliation{Keldysh Institute of Applied Mathematics, 
	4 Miusskaya sq., Moscow, 125047, Russia}

\begin{abstract}
The process of uniform supernovae explosions (SNe) is well investigated for all their types. However, observational data suggests that the SNe could be not spherically-symmetric. Modern multi-dimensional simulations of SNe demonstrate development of hydrodynamical instabilities during the explosion phase. But the configuration of a star and inhomogeneities prior to explosion could strongly affect how the SNe develops. In a number of papers on numerical modeling of pair-instability supernovae explosion (PISNe) considered the case when thermonuclear energy in the central region of a massive star is introduced by the series of several hot spots. It leads to the appearance of many fragments of hot matter behind the divergence shock wave. An observable manifestation of this may be the presence of peaks on light curves of gamma-ray burst associated with explosions of massive stars. 

The physical nature of such inhomogeneities is not evident and the number and size of spots is a conjecture. In this work, we study the possibility of formation of these inhomogeneities at the stage of the core-collapse \colFo{(CC)} in a massive star. To check this assumption, we chose analytic self-similar model of \colFo{CC} and investigated stability of solutions obtained from it with respect to small multidimensional perturbations. It shows there are no conditions where the collapse of a very massive star may remain stable, although, for a less massive star, it is possible. Using obtained relations, we found characteristic features of developing instability, thereby making it possible to estimate the amount and characteristic size of the inhomogeneities. 
\end{abstract}


\section{Introduction} \label{sec:intro}

\subsection{Overview} \label{subsec:overview}

Supernovae explosion \colFo{(SNe)} is a tremendous explosion that disrupts a star. \cols{All \colFo{SNe} are classified by their spectral characteristics and lightcurve features which show high diversity. The physical picture of \colFo{SNe} -- what triggers the loss of stability and what drives the explosion -- depends on many parameters -- the mass of a star, its chemical composition, rotation, etc.} 

\cols{Stars with masses $8M_\odot<M<40M_\odot$\colFo{, hereinafter referred to as massive stars,} evolve through stages of thermonuclear burning until the depletion of the fuel and formation of nickel-iron core. The fission of heavier elements requires energy, rather than release it. At some point internal pressure cannot support the core and it collapses triggering an explosion of a star called a core-collapse supernova \colFo{(CCSN), which is} usually observed as type-II supernova with presence of hydrogen lines.}

\cols{\cite{Fowler1964} discovered the importance of electron-positron pair formation and the role of neutrino losses in massive stars.} When the central temperature in the core reaches $T_c \simeq 2 \cdot 10^9\text{ K}$ intensive production of pairs occurs. Since \colIK{this pioneer} work numerous articles with detailed analyses have been done (\cite{Alsabti2015}). \cols{For massive stars neutrino emission plays a significant role. It influences the evolution of a star making the contribution to the energy loss and thus shortening lifetime. But also in \colFo{CCSN}, the role of neutrinos in the generation of the shock wave is crucial (\cite{Janka2007}). The density during \colFo{CC} can reach values higher than  $10^9\text{ g cm}^{-3}$, making inner shells non-transparent for neutrinos. Thus neutrinos put additional pressure helping to revive and accelerate the shock.}

However, the general picture of \colFo{explosions of stars with masses $M>40M_\odot$, hereinafter called massive stars,}  is \colFo{inherently} different from the less massive stars explosion. \cite{BisnovatyiKogan1967,Rakavy1967} \cols{showed that the $e^+e^-$ pair formation process can lead to unstable configuration of \colFo{very} massive stars at a relatively early stage of evolution (at the oxygen burning) when thermonuclear fuel is abundant. It can result in accelerating contraction and subsequent thermonuclear explosion that} disrupts a whole star. A new type of powerful and luminous supernova was called the pair-instability supernova (PISN).  This \cols{model was studied and confirmed} by numerous other works over the years: \cite{Fraley1968,Arnett1974,Wheeler1977,ElEid1983,Ober1983,Bond1984,Fryer01}.

Unlike massive stars $8M_\odot<M<40M_\odot$, the internal conditions of \cols{very massive stars} are different (\cite{Heger03}) mainly due to the fact that the central density is inversely proportional to the mass $\rho_c \sim T_c^3/ \sqrt{M}\sim 10^5 \text{ g cm}^{-3}$   and $T_c \sim 2 \cdot 10^9\text{ K} $. \cols{Thus, in very massive stars at the stages of oxygen burning the stellar matter is transparent for neutrinos and their role is limited only to the reduction of the internal energy and acceleration of the star's evolution. At the same time, due to the high temperature in the center of a star, the process of $e^+e^-$ pairs creation by gamma rays \colFo{gets} effective. Annihilating pairs again produce gamma rays and a few amount of neutrinos. Since the cross-sections of pair-creation and annihilation are finite, gamma rays and $e^+e^-$ pairs reach thermal equilibrium. Thus the energy of photons transforms to the rest mass energy of the pairs, altering the equation of state \colFo{(EoS)} and reducing pressure. }\cols{In order to restore hydrostatic equilibrium, the star contracts and the temperature of \colFo{stellar} subsoil increases, increasing the equilibrium number of $e^+e^-$ pairs. This leads to additional decrease of pressure, compression of the \colFo{stellar} nucleus and even greater heating, which results in the formation of even more $e^+e^-$ pairs. It triggers an avalanche-like process of collapse of the star. Due to profusion of nuclear fuel, when the temperature in the center reaches high enough values, the rate of nuclear reactions rapidly increases by a few orders of magnitude, which could lead to explosive nuclear burning and reverse the collapse.} Almost all 1D codes reach this schematic picture (\cite{Paxton2013}) which shows that these physical processes are well understood.

\smallskip\smallskip\smallskip\smallskip\smallskip
\subsection{\colFo{Sphericity of core-collapse}}

In order to have a detailed picture of the hydrodynamic explosion and then, to investigate the mixing, multidimensional codes are needed. Nowadays, using high performance computing, this field proposes a laboratory view of such a tremendous event.
However, the ignition itself still remains a puzzling issue. Where does the ignition take place and how does it develop? 
One of possible explanations is the following: nuclear burning in the center of a star might develop large-scale convection (\cite{Bandiera1984,Arnett2011}) which spoils spherical symmetry of the system. Inhomogeneities in temperature and density could cause ignition spots to occur in the core. Following this idea, it has been possible to propose asymmetrical explosion using multicore ignition and to show that an asymmetrical explosion will produce a completely different hydrodynamic picture of the explosion. {\cols{Results of other investigations indicate \colFo{the development of} Rayleigh-Taylor instability during the explosion (\cite{Chen2011,Chen14,Gilmer2017}), which also could spoil the spherical scenario}.}

\colFo{Several investigations of CC stability were concentrated on accretion without taking into account the self-gravity of the collapsing gas (\cite{Kovalenko1998,Blondin2003}). The \colFo{CC} stability of self-gravitating polytropic gas was considered by \cite{Lai2000}. They showed that nonspherical perturbations grow in the supersonic region approaching to $\Delta\rho/\rho\sim r^{-l}$ law, where $l$ is the angular degree of a perturbation. The future consideration (\cite{Cao2009,Lou2012}) with general polytropic \colFo{EoS} showed the importance of perturbation modes which are driven by buoyancy and trapped deep inside the core. Although such considerations do not allow one to identify a dominant mode of instability, it has important physical consequences. Thus, in the context of the \colFo{CC} of massive stars with a compact remnant, $l=1$ mode may be responsible for initial kicks of the central compact object and its high speed; $l=2$ may contribute to formation of a binary system of compact objects; higher modes may break the whole core and prevent the formation of compact remnant.}

\colFo{Numerical simulations of an oxygen shell over a contracting core within $18M_\odot$ star shows (\cite{Mller2016}) the emergence of $l=3$ and $l=4$ modes which latter gets replaced by $l=2$ mode. The latter investigations (\cite{Vartanyan18,Bollig2021}) demonstrate that perturbations in such a shell determine the asymmetry of the mass ejection, which has a bipolar structure. As it was shown by \cite{Powell2021} with numerical simulations, \colFo{CC} of $85M_\odot$ and $100M_\odot$ stars is also nonspherical because of  standing accretion shock instability (\cite{Blondin2006,Foglizzo2007}). Thereby, in the stars of the above masses, instability during collapse can develop. The question of the possibility of collapse instability development in PISN progenitors remains open.}

\subsection{\colFo{PISNe -- GRB connection}}

In previous articles (\cite{Chardonnet2010,Baranov2013,Chardonnet2015}) we proposed a new scenario of PISN explosion with fragmentation. \colIK{In the explosion numerical simulations the initial conditions were chosen as a set of several randomly distributed hot spots in the core. Such multicore ignition leads to the asymmetrical explosion and a complex structure of divergence shock wave. The assumption about the nonuniform explosion} \cols{was made in order} to explain observational signatures of SN  which are different from a SN type II-P and to obtain spectrum and light curves similar to the one of gamma-ray bursts (GRBs) during the prompt \cols{phase}. Thus, the engine of GRBs occurs to be related to nuclear energy of the explosion of a PISN {\cols{(see also \cite{Fryer01,Kozyreva15,Gilmer2017,Chatzopoulos19,Roy2021})}}. 
 In order to conceptualize the phenomena, we propose to tackle this problem from another perspective. 

 \cols{In this work we investigate the idea} that temperature and density inhomogeneities\colIK{, which also were used as the initial conditions for numerical simulations,} \cols{could be formed} already at the stage of \colFo{CC}, which further leads to a non-uniform supernova explosion. And we make an attempt to extract some physical quantities which contribute to the development of inhomogeneities during the collapse, using analytical approach. It may be done by reanalyzing this problem from the stability point of view, using analytical methods similar to the methods of studying the converging shock wave problem. \colFo{Our} goal is to propose new physical solutions for the initial configurations of this complex and multifaceted problem to be used in multidimensional codes afterwards. Such analytical approach is \cols{especially simple in the case of \cols{the very massive stars}}, where the role of neutrinos can be simplified as an energy loss function  in the hydrodynamic equations. \col{We believe that such analytical investigation may point on a way for multidimensional numerical simulations \colIK{and motivate the future investigations in this area}. }

\subsection{\col{Strategy}} \label{subsec:stratetgy}

\cols{We can try to approach the question of how a \colFo{PISN} explodes, not only by numerical simulations, but using analytical methods in order to find possible regularities.} \cols{\col{The role of instabilities in the development of the explosion is an exciting part of the problem.} As  we have mentioned in the Section~\ref{subsec:overview}, in very massive stars the $e^+ e^{-} $ pairs are produced copiously affecting the \colFo{EoS}.} \col{From the thermodynamic point of view, the $e^+ e^{-} $ pairs creation reduces the adiabatic index $\gamma= (\partial\ln{p}/\partial \ln{\rho})_S$. \cols{Pair-instability develops when substantial regions of a star reach conditions} where $\gamma$ is less than the critical value ${4}/{3}$ (see e.g.\@ \cite{Cox1968,Nadyozhin1995}). If enough mass of the star enters in those regions, it becomes unstable, \cols{contracts and explodes as PISN. Thus, a important step towards understanding the dynamics of explosion is to explore the initial contraction phase. Our approach is \colFo{to} investigate possible departure from a spherical contraction} using the basic stellar physics equations. Several steps are required for such description of the instabilities. Our method will follow the self-similar collapse computation of \cite{Nadezhin1969}. In spite of some differences to our problem, it  is not an exaggeration to argue that this method is very useful to tackle the solution inside a certain domain of parameters. Even if the main similarities are the basic equations and the neutrino losses, a suitable variation of the thermodynamic index is necessary. Once the solutions will be derived analytically, we will study the instabilities and derive some criteria. }

\section{Self-similar solutions of core-collapse} \label{sec:nadez}

\subsection{Equations and boundary conditions} \label{subsec:equsAndBound}

In order to investigate the problem, we chose and slightly modified the well-known model of \colFo{CC} through energy losses by neutrino emission, created by \cite{Nadezhin1969}. In this model, the \colFo{CC} is described by a self-similar solution of hydrodynamic equation using the general method of self-similar solutions (\cite{Zeldovich1967b}). \col{The $e^+e^-$ pairs creation reduces adiabatic index $\gamma$ to values less than 4/3. If a significant part of a star reaches conditions where $\gamma < 4/3$, the star becomes ''soft'' and dynamically unstable. Therefore, in our consideration we had to include the possibility of changing $\gamma$ as a function of radius and time.}

It must be noted that behavior of spherical converging shock waves is well investigated using the same method since it is related to the problem of ignition of the nuclear fusion reaction with lasers (\cite{badziak}). As it was shown by \cite{Brushlinskii1982,murakami} such shock waves are unstable if they propagate in medium with constant density. We do not know what effect on stability is due to the inclusion of self-gravity, radiation cooling and a heterogeneous medium which take place inside a star. Here we \col{describe the receipt of such a solution} in order to proceed in Section~\ref{sec:stabInv} to the analysis of \colFo{small} perturbations of this system.

\colFo{The equations} of gas dynamics, taking into account the radial self-gravity and the energy losses by neutrino radiation, written in customary notation as follows:
\begin{align}
& \frac{\partial \rho}{\partial t} + \nabla\mathbf{\cdot} \rho \mathbf{v} = 0,  \label{parSysR} \\
& \frac{\partial \mathbf{v}}{\partial t} + (\mathbf{v}\mathbf{\cdot} \nabla)\mathbf{v} = -\frac{1}{\rho} \nabla p - \frac{GM}{r^2} \mathbf{\hat{r}},  \label{parSysV} \\
& \frac{\partial M}{\partial r} = 4 \pi  \rho r^2, \\
& \frac{\partial \epsilon}{\partial t} + \mathbf{v}\mathbf{\cdot} \nabla \epsilon -\frac{p}{\rho^2} \left( \frac{\partial \rho}{\partial t}  + \mathbf{v}\mathbf{\cdot} \nabla \rho  \right) + Q_\nu =0, \label{parSysP}
\end{align}
where \colFo{the EoS has the form $p = b \rho^\sigma T^\eta$ ($a,b>0$),} $\mathbf{\hat{r}}$ -- radial unit vector, the internal energy has the form: $\epsilon = (\gamma-1)^{-1} p \rho^{-1}$ \colFo{and $Q_\nu = a \rho^s T^n$ is rate of energy loss}. The parameters $s$, $n$, $\sigma$, $\eta$ and the adiabatic index $\gamma$ are supposed to be free parameters of the model, but there is relation following from the thermodynamic consideration: $\gamma = (\sigma - \eta)/(1-\eta)$ for $\eta\neq 1$. When $\eta=\sigma =1$ all usual values of $\gamma$ are possible. 

The system (\ref{parSysR})-(\ref{parSysP}) may be reduced to a system of ordinary differential equation using the general method of self-similar solutions deriving (\cite{Zeldovich1967b}). It needs to introduce a new self-similar variable $x=r/R(t)$, where the characteristic length scale is $R(t)=At^\alpha$, $A$ and $\alpha$ are constants which values will be set later. Then the unknown functions may be found in the form: $\rho={\rho_\text{in}}(t) g(x)$, $p={\rho_\text{in}} (t)\dot{R}^2(t)h(x)$, $\mathbf{v}=\dot{R}(t)f(x)\mathbf{\hat{r}}$ and $M=4\pi{\rho_\text{in}}(t)R^3(t)m(x)$, where $g$, $h$, $f$ and $m$ are dimensionless functions, the dot denotes the time derivative. \col{Unlike the work of \cite{Nadezhin1969} we also let $\gamma$ be a function depending on the self-similar variable as $\gamma = \Gamma(x)$. }

In such problems \colFo{it is usually assumed} that $t=0$ is the moment of focusing, i.e.\@ the moment, when the density in the center becomes infinity. Therefore, the transformation $t\rightarrow -t$ is supposed to be done, so the self-similar variable looks like $x=r/A(-t)^\alpha$. It means at $t=-\infty$ the collapse starts and it finishes at $t=0$. Thus, for any point lying outside the center $r=0$ we have $x\rightarrow \infty$ while $t\rightarrow 0$ and $x\rightarrow 0$ when $t\rightarrow -\infty$. Hereinafter this transformation is supposed to be done.

After such substitutions the following system might be got:
\begin{align}
& f' + (f+x)\frac{g'}{g} + \frac{2f}{x} + \frac 2 \alpha = 0, \label{ssR} \\
& (f+x)f' + \frac{1-\alpha}{\alpha} f = -\frac{(g\tau)'}{g} - \frac{m}{x^2}, \label{ssV} \\
& m' = x^2 g, \label{ssM} \\
&(f+x)\frac{g'}{g}\tau - \frac{f+x}{\Gamma-1}\tau' \col{+ \frac{f+x}{(\Gamma-1)^2}\tau \Gamma'} +\nonumber \\  
 & + \frac 2 \alpha \frac{\alpha+\Gamma-2}{\Gamma-1}\tau - g^\nu \tau^\chi = 0, \label{ssE} 
\end{align}
if we set special values of ${\rho_\text{in}}$, $A$, $\alpha$:
\begin{align}
&{\rho_\text{in}} = \frac{\alpha^2}{4\pi G t^2}, \\
&A = \left( \frac{(4\pi G)^\nu b^\chi }{a\alpha^{2(\nu+\chi)-3}} \right)^{1/{2(\chi-1)}},\label{AspT} \\
&\alpha = \frac{\chi +\nu -3/2}{\chi-1} \label{ALPHAspT} ,
\end{align}
where the prime denotes the $x$ derivative, $\nu=s-(\sigma-1)n/\eta$, $\chi=n/\eta$ and $\tau = h/g$. It is possible to lower the order of the system (\ref{ssR})-(\ref{ssE}) by introducing so-called mass integral (\cite{Sedov1959}), which is the consequence of (\ref{ssV})(\ref{ssM}):
\begin{equation}
\left( 3 - \frac 2 \alpha \right) m = x^2 g (f+x).
\label{massI}
\end{equation}
So, if $\chi\ne 1$ and $\alpha\ne 2/3$ we have the system of three ordinary differential equations for three values: $g$, $f$, $\tau$. 

There is one obvious boundary condition $f(0)=0$. As it was shown, at $x=0$ the following relation takes place:
\begin{equation}
g^\nu(0)\tau^\chi(0) = \frac{2}{\alpha}\frac{\col{\Gamma(0)}+\alpha-2}{\col{\Gamma(0)}-1}\tau(0).
\end{equation}
Thus only one \cols{unknown} parameter $g(0)$ \colFo{remains}. The system (\ref{ssR})(\ref{ssV})(\ref{ssE}) with substitution (\ref{massI}) has a singular point $x_s$, where the denominator $H$ of an explicit expression $df/dx = F/H$ becomes zero. From the condition that the numerator $F$ turns to zero at the singular point, it is possible to find the unknown $g(0)$. \cite{Nadezhin1969} used the following procedure. Firstly an arbitrary point $x_c$, which is located between the boundary and the singular point $0<x_c<x_s$, is taken. Then, the system is integrated from $x=0$ to $x=x_c$ with a test value of $g(0)$ and from point $x_s$ to $x_c$ with trial values of $x_s$ and $f(x_s)$. Values $g(x_s)$ and $\tau(x_s)$ may be found from requirement $F=H=0$ at $x_s$. At the matching point $x_c$ all three functions $g$, $f$, $\tau$ must be continuous. The three unknown parameters $g(0)$, $x_s$, $f(x_s)$ may be defined from this requirement. The system is then solved on the interval $x_s<x<+\infty$ with the known conditions at $x=x_s$ and the problem is considered to be solved. 

From mathematical point of view, the problem of finding a correct value of $g(0)$ may be reduced to finding a zero of the function $\{ g(0),x_s,f(x_s) \} \rightarrow \{ g(x_c^-)-g(x_c^+), f(x_c^-)-f(x_c^+), \tau(x_c^-)-\tau(x_c^+) \}$, calculated numerically. The parameters $s$, $n$ determining rate of energy loss, $\sigma$, $\eta$ determining the \colFo{EoS} and adiabatic index \col{$\Gamma(x)$ are set manually to get the model describing a process of interest}.
The full investigation of existence and uniqueness of a solution of the problem for different values of free parameters is not in frames of our work. The limit of applicability of this model is discussed by \cite{Murzina1991} and we checked the results in the Section~\ref{subsec:phyRes} according to this.

\subsection{Found solutions} \label{subsec:sols}

\begin{figure}
\includegraphics[width=1\linewidth]{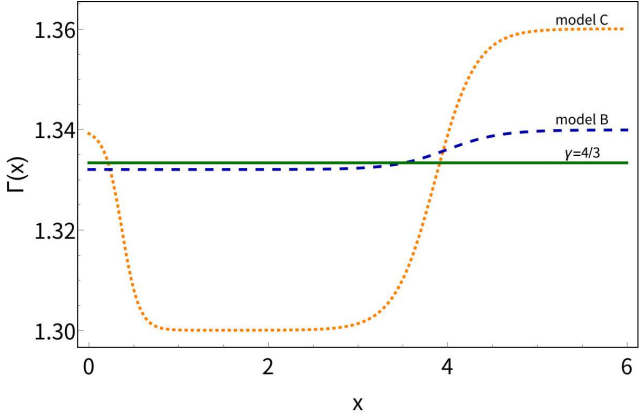}
\caption{Chosen $\Gamma(x)$ for model~B (dashed line) and model~C (dotted).}
\label{fig:gammas}
\end{figure}

\begin{deluxetable}{cccccccc}
\tablecaption{Found conditions for the sets of parameters \label{tab:params}}
\tablewidth{0pt}
\tablehead{ 
 \vspace{-0.4cm} & \colhead{Neutrino} & & & & & & \\ 
\colhead{\vspace{-0.4cm}  Model} & \colhead{emission} &  \colhead{$\Gamma(x)$} & \colhead{$\nu$} &  \colhead{$\chi$} & $g(0)$ & \colhead{$x_s$}  & \colhead{$f(x_s)$} \\   & \colhead{mechanism} & \colhead{} & \colhead{} & \colhead{} & &  & }   
\startdata
A & Urca & $5/3$ & $0$ & $6$ & $4.825$  & $2.644$ & $-1.348$ \\
B &      & $\Gamma_\text{B}(x)$ & $0$ & $3$ & $52.918$  & $1.408$ & $-0.707$ \vspace{-0.18cm}\\
\vspace{-0.18cm}  & Pair &       &     &     &           &          &            \\
C &      & $\Gamma_\text{C}(x)$ & $0$ & $3$ & $20.657$ & $1.577$ & $-0.900$  \\
\enddata
\tablecomments{ 
$\Gamma_\text{B}(x) = 1.34 - 0.0039 \tanh(6.7 - 1.68 x)$,\\
$\Gamma_\text{C}(x) = 1.35 + 0.02 \tanh(1.92 - 5.2 x) - 0.03 \tanh(8.5 - 2.2 x)$. \\
See Figure \ref{fig:gammas}.  }
\end{deluxetable}

\begin{figure}
   \centering
\begin{tabular}{c}
\includegraphics[width=\linewidth, height=\linewidth, keepaspectratio]{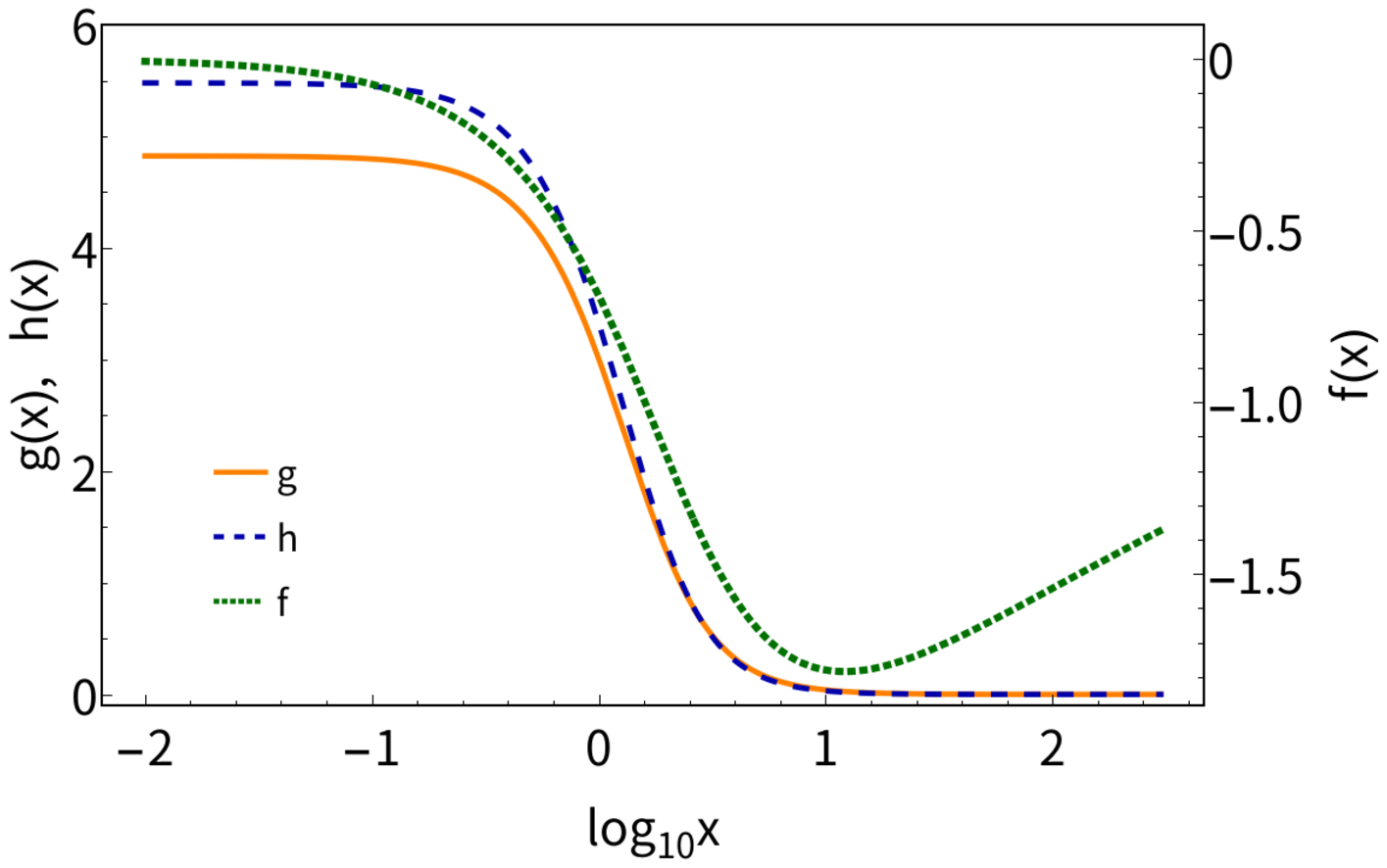}\\
\includegraphics[width=\linewidth, height=\linewidth, keepaspectratio]{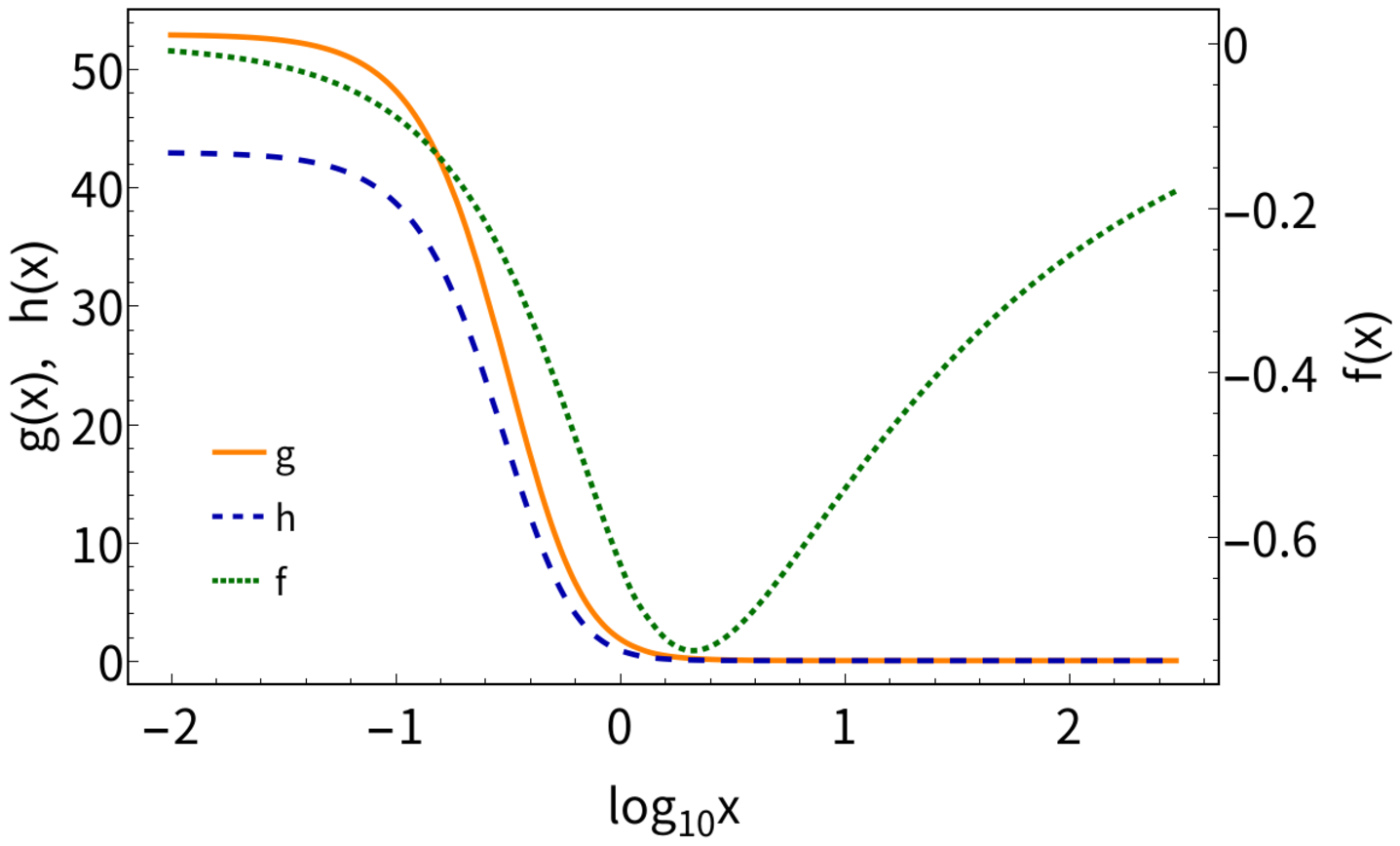}\\
\includegraphics[width=\linewidth, height=\linewidth, keepaspectratio]{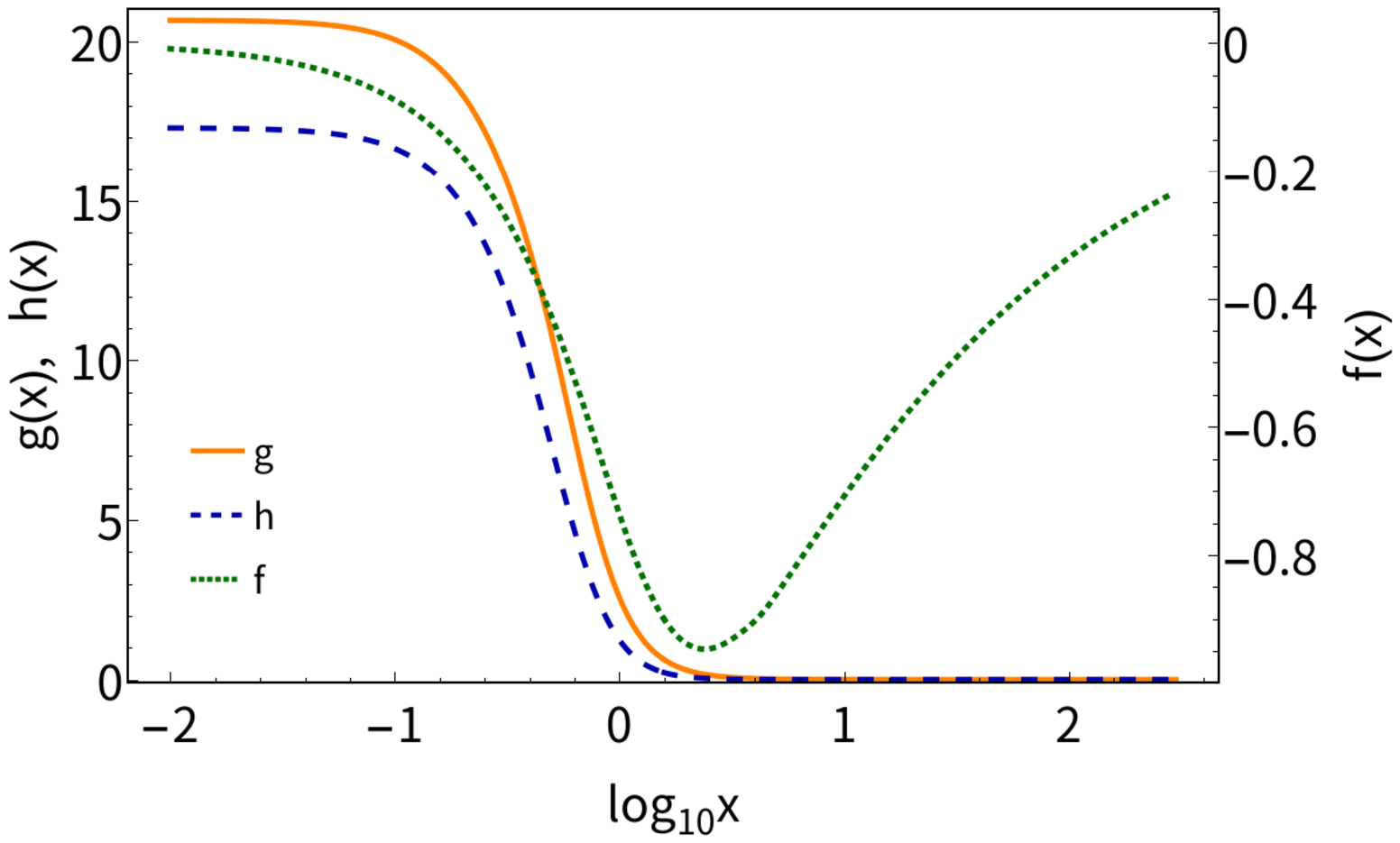}\\
\end{tabular}
\caption{The solution of unperturbed system (\ref{ssR})(\ref{ssV})(\ref{ssE})(\ref{massI}) for models~A (top panel), B (middle) and C (bottom): dimensionless density is the solid line, dimensionless pressure and velocity are the dashed and dotted lines correspondingly.}
\label{fig:SolABC}
\end{figure}

\begin{figure*}
\begin{minipage}{0.32\linewidth}
\includegraphics[width=1\linewidth]{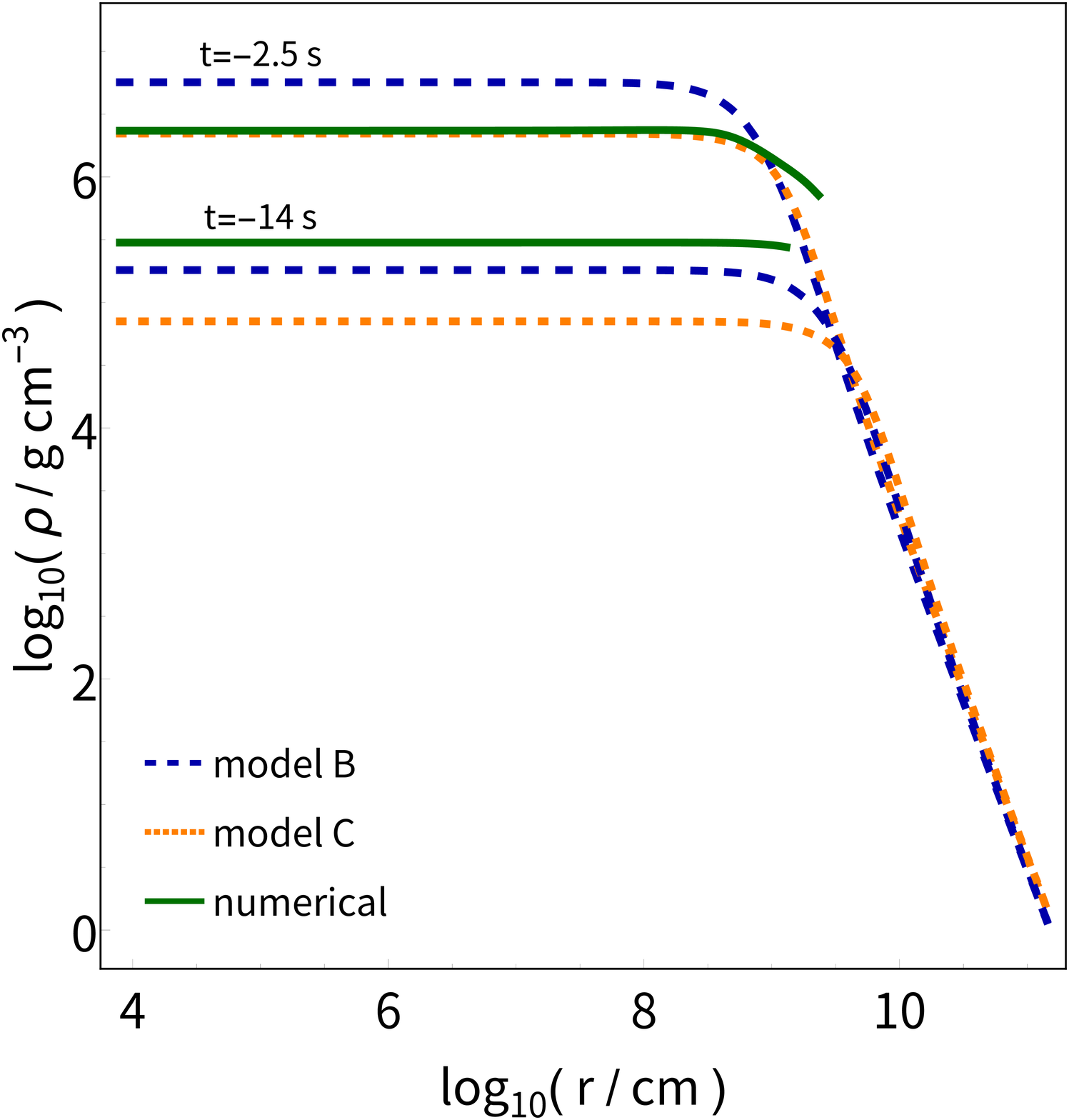}
\end{minipage}
\hfill
\begin{minipage}{0.32\linewidth}
\includegraphics[width=1\linewidth]{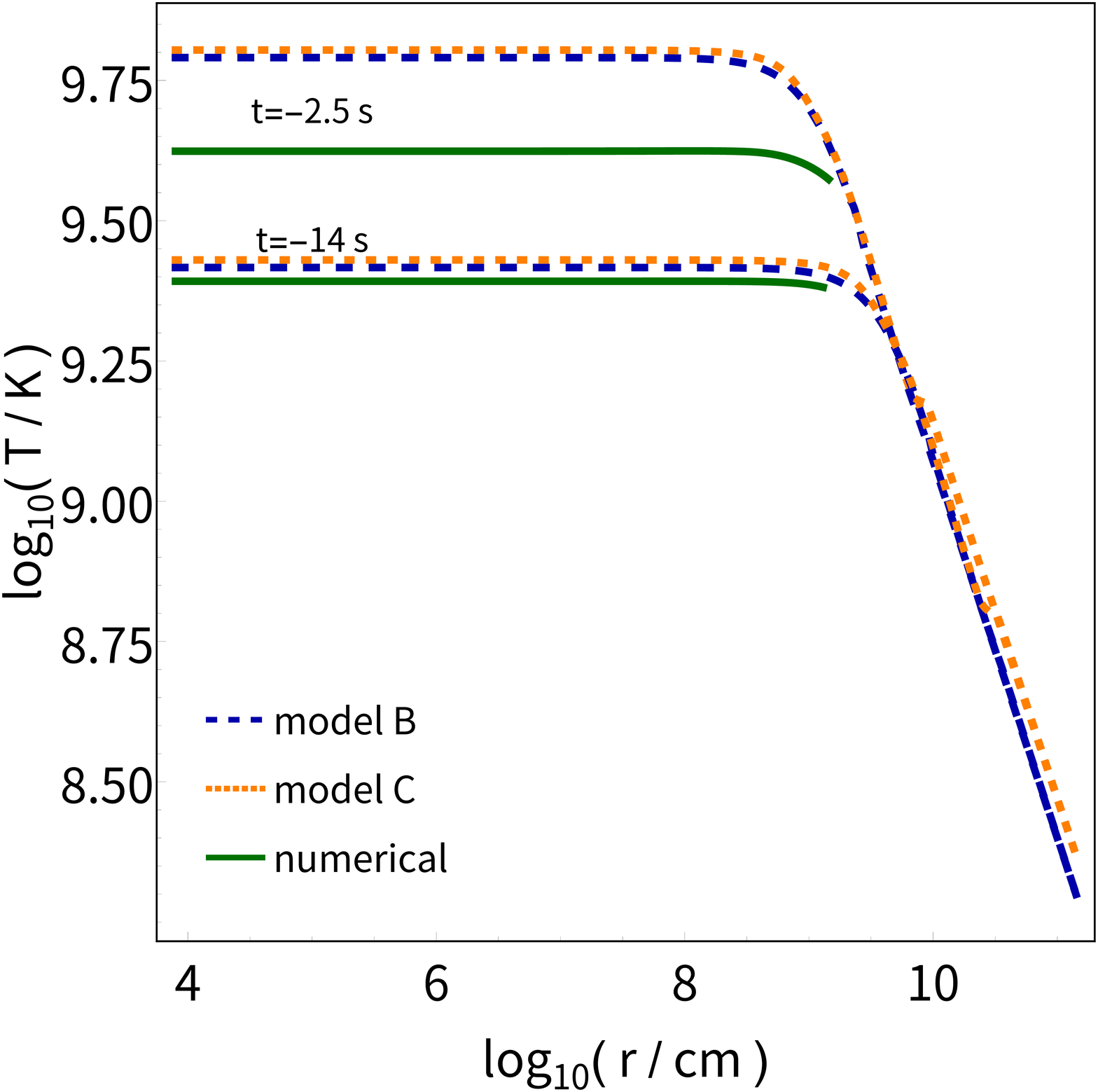}
\end{minipage}
\hfill
\begin{minipage}{0.32\linewidth}
\includegraphics[width=1\linewidth]{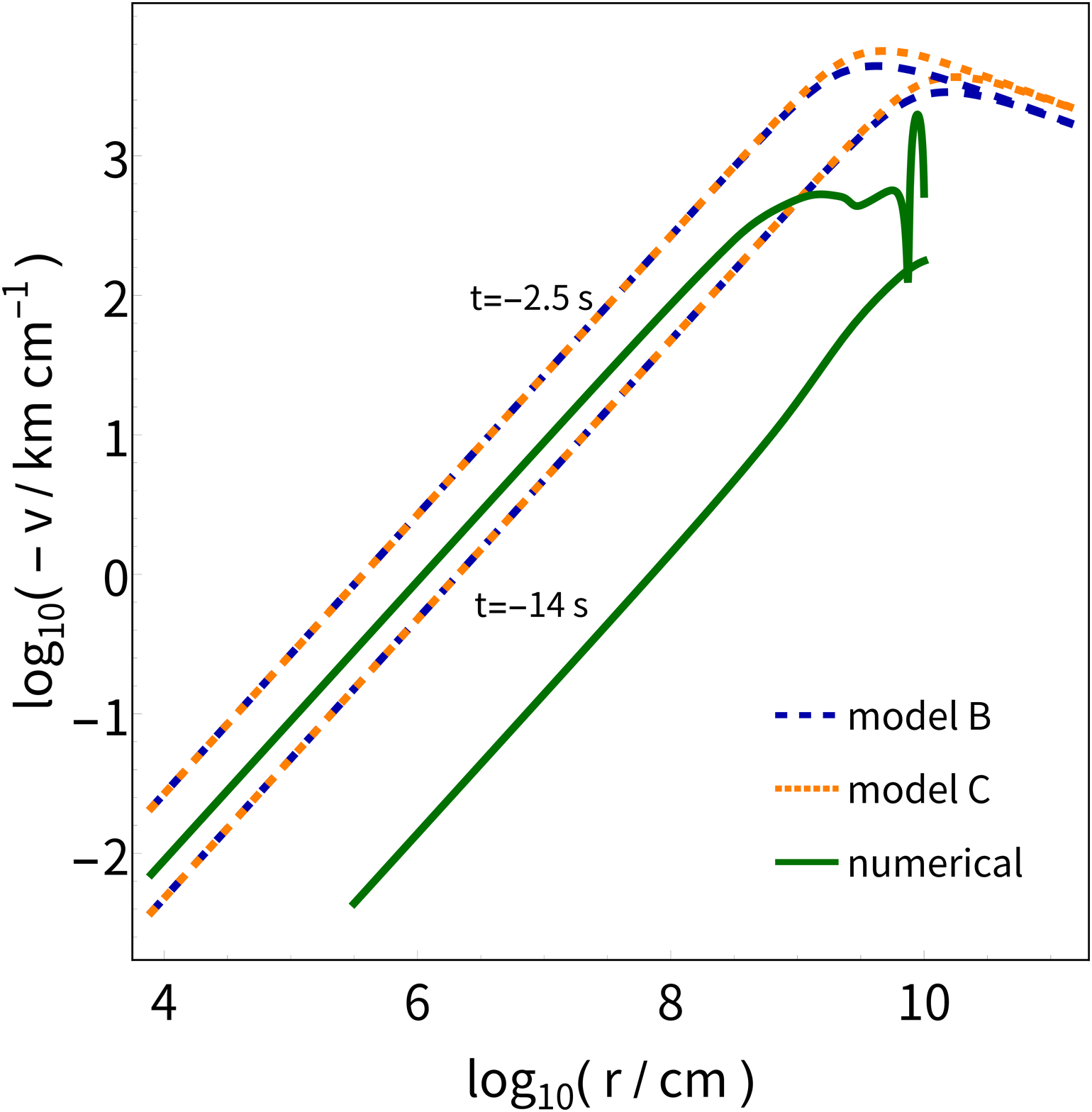}
\end{minipage}
\caption{From left to right: distributions of density, temperature and velocity at $t=-2.5\text{ s}$ and $t=-14\text{ s}$ for models~B (dashed line), C (dotted) \colIK{and 1D numerical modeling \colFo{according to \cite{Baranov2013}} (solid)}.}
\label{fig:physBC}
\end{figure*}

To find the conditions of passing the special point, we were looking for a minimum of the function $U(g(0),x_s,f(x_s))= (g(x_s^-)-g(x_s^+))^2 + (f(x_s^-)-f(x_s^+))^2 + (\tau(x_s^-)-\tau(x_s^+))^2$ using a method developed by \cite{Brent1973}, because it does not require the calculation of derivatives of a minimized function. The differential equations were solved using the predictor-corrector method. For all the solutions found, the value of the function $U$ at the minimum does not exceed~$10^{-10}$. Under such consideration, it is impossible to state unequivocally that a solution was not found due to the fact that it does not exist -- this is a drawback of the method used. Therefore, we restricted ourselves by considering only those cases for which we managed to find a solution and which are related to physical processes inside a star.

\col{Using the method described above, we reproduced the solution given by \cite{Nadezhin1969}, describing the \colFo{CC} via Urca-process with ideal \colFo{EoS} ($\sigma=\eta=1$) and constant $\Gamma=5/3$, which is featured below as model~A.}

\col{ To make the model describe \colFo{CC} of PISN progenitor, we also chose ideal \colFo{EoS}, because apparently in a rough approximation degeneracy of the stellar matter can be neglected. We considered two cases of $\Gamma(x)$ -- in the first (model~B) it keeps $>4/3$ in the outer part of a core and cross $4/3$ on the way to the center. In the second case (model~C), it has more amplitude of changing and $>4/3$ in outer part and near the center (see Figure~\ref{fig:gammas}).}
\cols{The shape of $\Gamma$ of the model~C was taken near to the work of \cite{Gilmer2017}, which shows a plot of $\gamma$ at the moment close to an explosion. At the same time, we had to underestimate the minimum value of $\Gamma$ to almost $1.3$, because for large values it was not possible to construct a solution according to our method. Model~B is aimed to reveal the role of a particular form of $\Gamma(x)$, so it was chosen to have $\Gamma(x)$ with a different behavior and a much smaller amplitude of change. As it will be seen in Section~\ref{sec:Analysresult}, a reasonable form of the $\Gamma(x)$ dependence does not make a significant contribution to the collapse stability, which allows us to choose $\Gamma(x)$ behavior in a wide range and still get meaningful results of stability investigation.} 

 \col{Since $\gamma$ can depend only on the self-similar variable, its shape is the same for any moment in time -- it may only squeeze while $t$ decreases, i.e.\@ there is no moment when the $\gamma$ becomes less than $4/3$ being larger before. We believe that our description becomes usable starting from the moment when some part of a star has already passed into the state with $\gamma <4/3$. \colFo{According to works mentioned in Section~\ref{sec:intro} it happens no later than approximately $20\text{ s}$ before the moment of maximum compression.} }

\col{ Another important step forwards the construction of an analytical solution was to describe the neutrino losses. Here, we used results of \cite{schinder}. The authors} proposed a fitting formula, describing the energy loss by neutrino emission because of plasma, pair and photo processes as a function of density and temperature; in our definitions they calculated $\rho Q_\nu$ value. Of course, the complexity of their formula does not allow the construction of a self-similar solution with its using. Almost all quantities presented in the formula have the same order for interesting \colFo{for us} parameters, because of what it is impossible to distinguish the dominant term; so we acted differently. With typical conditions inside a very massive star $\rho_c \sim 10^5 \text{ g cm}^{3}$ and $T_c \sim 10^9 \text{ K}$ we may approximately set $Q_\nu\simeq [9\cdot 10^{-16}\text{ cm}^2\text{s}^{-3}\text{ K}^{-3}] T^3$ for pair neutrino emission. Such simplification provides values of $Q_\nu$ consistent with the results of \cite{schinder} in a neighborhood of the typical parameters \col{inside a massive star. Thus for models~B~and~C $s=0$, $n=3$ were chosen.}

\col{Parameters of models and found conditions of passing the special points are summarized in Table~\ref{tab:params}. The found solutions are presented in Figure~\ref{fig:SolABC}. It can be seen that model~B shows twice more compression than model~C. Wherein a slightly higher speed is achieved in model~C. It seems to be related with presence of $\Gamma>4/3$ region near the center for model~C, due to which the core of a star shows greater ''rigidity''.}

\colIK{We also performed 1D numerical simulations of \colFo{CC} of a star with $110 M_\odot$ core mass using a code described by \cite{Baranov2013}, \colFo{the computational domain of which extends} up to $10^{10}\text{ cm}$. This code takes into account the exact \colFo{EoS} and the neutrino loss law. It is also possible to make \colFo{comparison} with the already existing calculations provided by \cite{Chatzopoulos2013}. 
	To compare this numerical result with our models \colFo{we choose two time moments first of which corresponds to a moment of maximum compression in the numerical simulation. Then, in our model, such a moment in time was chosen so that the density of model~C coincided with the numerically calculated one. For models~B~and~C it is $t=-2.5\text{ s}$ with $A=10^9 \text{ cm s}^{-0.75}$, which corresponds to an oxygen core. The second moment was chosen $11.5\text{ s}$ apart from the first for all models.} As can be seen from Figure~\ref{fig:physBC}, models~B~and~C quite accurately reproduce densities and temperatures both our numerical calculation and one presented by \cite{Chatzopoulos2013}, although they do not show some features \colFo{like presence of inflection points taking place in the work of \cite{Chatzopoulos2013}}, which should not be expected from a simple analytical consideration. \colFo{Wherein, the compression in our model is faster and velocities is higher, which can be associated with both inaccurate determination of $A$ and model dependencies $\Gamma(x)$.}}

\begin{figure}
	\centering
	\begin{tabular}{c}
		\includegraphics[width=\linewidth, height=\linewidth, keepaspectratio]{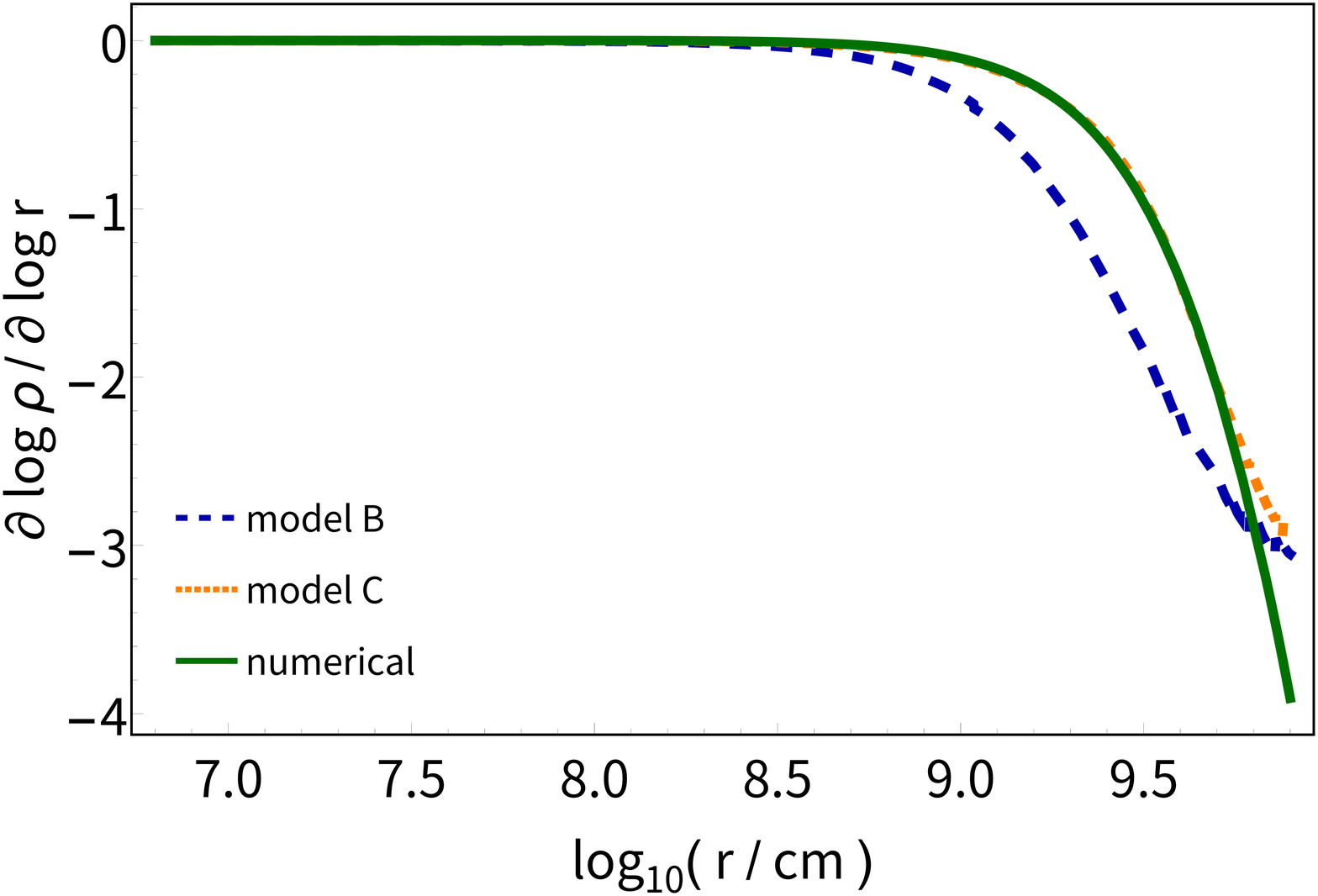}\\
		\includegraphics[width=\linewidth, height=\linewidth, keepaspectratio]{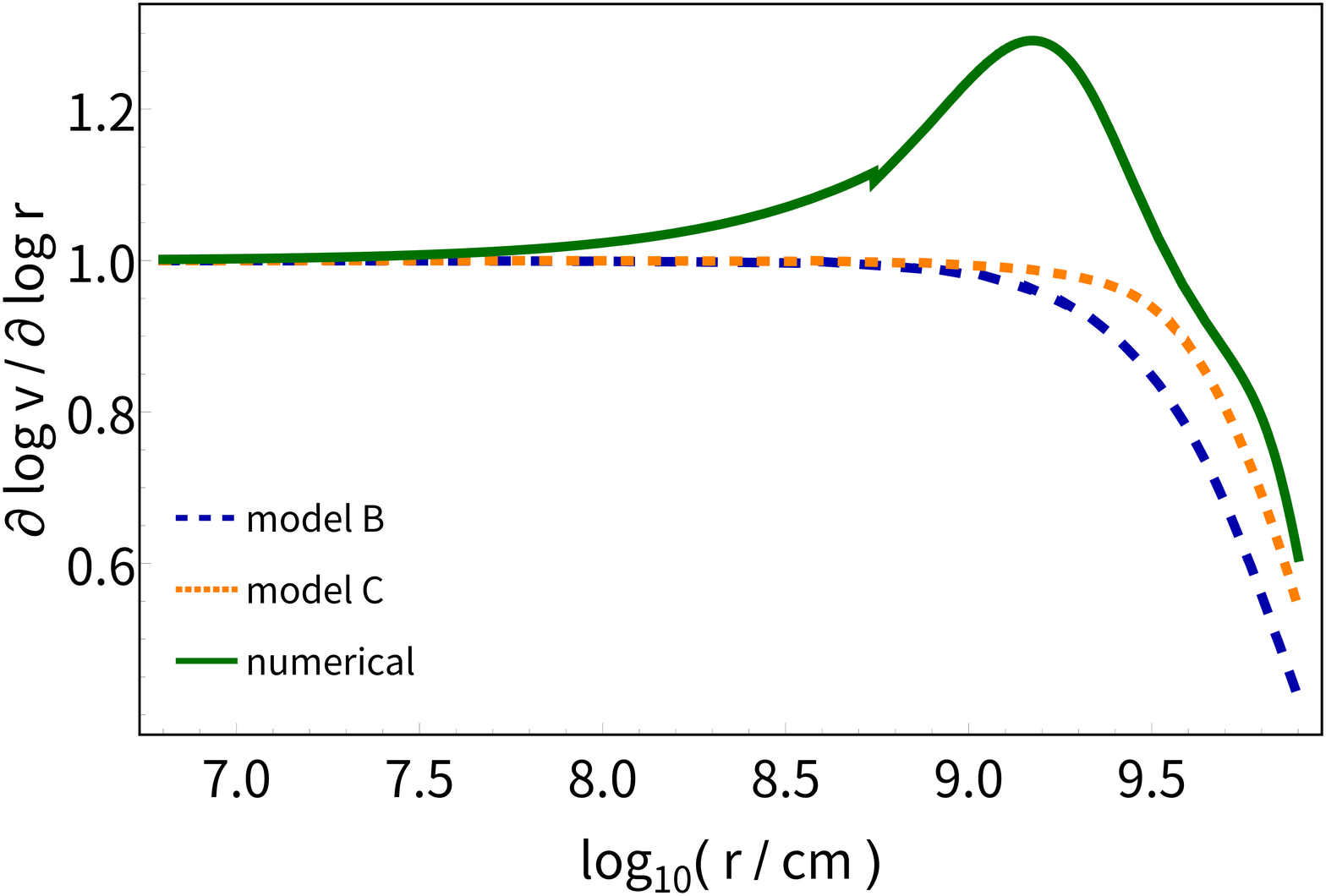}\\
	\end{tabular}
	\caption{Dimensionless numbers $({r}/{\rho})({\partial \rho}/{\partial r})$ (top panel) and $({r}/{v})({\partial v}/{\partial r})$ (bottom) characterizing the changing rate of the corresponding values. Models~B (dashed line), C (dotted) and 1D numerical modeling according to \cite{Baranov2013} (solid) were taken at the moment $t=-14\text{ s}$.} 
	\label{fig:AT}
\end{figure}
\colFo{The formation of instabilities is significantly affected by the gradients of unperturbed quantities. Therefore, in order to assess how suitable the obtained solution is for studying it for stability, we calculated the value  $({r}/{\rho})({\partial \rho}/{\partial r}) = {\partial \log\rho}/{\partial \log r}$, which is an analogue of the Atwood number for a uniformly distributed density. Since the velocity gradient can also affect stability, a similar value $({r}/{v})({\partial v}/{\partial r})$ was considered. We considered moment $t=-14\text{ s}$ because it is far enough from moment of focusing. As can be seen in Figure~\ref{fig:AT}, for both quantities the difference between our models and numerical calculations does not exceed $0.5<1$. Therefore,} presuming that the obtained solutions accurately enough describe the process of homogeneous \colFo{CC}, we examined them for stability.

\section{Stability investigation} \label{sec:stabInv}

\subsection{Linearized equations for perturbations} \label{subsec:equs}

Let us consider small perturbations of the self-similar solutions. For this we solved the system of ideal hydrodynamic equations (\ref{parSysR})-(\ref{parSysP}), finding a solution in the form of $\rho(\mathbf{r},t) = \rho_0(r,t) + \rho_1(\mathbf{r},t)$, $p(\mathbf{r},t) = p_0(r,t) + p_1(\mathbf{r},t)$, $M(\mathbf{r},t) = M_0(r,t) + M_1(\mathbf{r},t)$, $\mathbf{v}(\mathbf{r},t) = v_0(r,t)\mathbf{\hat{r}} + \mathbf{v}_1(\mathbf{r},t)$, where the index $0$ denotes the above solution, index $1$ corresponds to the small perturbation which are supposed to be much less than the values of the unperturbed solution, therefore in the equations only linear terms should be left. \col{Due to the smallness of the perturbations the functional behavior of $\gamma$ should not change; therefore we did not consider its perturbations.} Following by  \cite{Brushlinskii1982,murakami} the solution may be found in the self-similar form:
\begin{align}
& \rho_1 = {\rho_\text{in}}(t) \sum_{l=0}^\infty \sum_{m=0}^l \left(\frac{t}{\tau}\right)^{\lambda_{lm}} \Phi_{lm}(x) Y_l^m(\theta,\phi), \label{subsRho} \\
& p_1 = {\rho_\text{in}}(t) \dot{R}^2(t) \sum_{l=0}^\infty \sum_{m=0}^l \left(\frac{t}{\tau}\right)^{\lambda_{lm}} \Omega_{lm}(x) Y_l^m(\theta,\phi),\label{subsP} \\
& M_1 = 4\pi{\rho_\text{in}}(t) R^3(t)\sum_{l=0}^\infty \sum_{m=0}^l \left(\frac{t}{\tau}\right)^{\lambda_{lm}} \Lambda_{lm}(x) Y_l^m(\theta,\phi), \label{subsM} \\
& v_{1r} = \dot{R}(t) \sum_{l=0}^\infty \sum_{m=0}^l \left(\frac{t}{\tau}\right)^{\lambda_{lm}} \Upsilon_{lm}(x) Y_l^m(\theta,\phi), \label{subsVr} \\
& v_{1\theta} = \dot{R}(t)\sum_{l=0}^\infty \sum_{m=0}^l \left(\frac{t}{\tau}\right)^{\lambda_{lm}} \Psi_{lm}(x) \frac{\partial Y_l^m}{\partial \theta}(\theta,\phi), \label{subsVt} \\
& v_{1\phi} = \dot{R}(t)\sum_{l=0}^\infty \sum_{m=0}^l \left(\frac{t}{\tau}\right)^{\lambda_{lm}} \Psi_{lm}(x)\frac{1}{\sin\theta} \frac{\partial Y_l^m}{\partial \phi}(\theta,\phi), \label{subsVp}
\end{align}
where ${\rho_\text{in}}(t)$, $R(t)$ and $x=r/R(t)$ are the same functions, being used in the original solution, $Y_l^m(\theta,\phi)$ are spherical harmonics, the capital Greek letters denote the unknown amplitudes of perturbations. $\tau<0$ is a fixed time of a measurement of the amplitudes, which disappears from the following equations. Unknown ${\lambda_{lm}}$ are grown rates to be found as eigenvalues for each mode. They may also be complex, the imaginary part corresponds to oscillation of the perturbations. Also, it is necessary to keep in mind the $t\rightarrow -t$ transformations. Such consideration does not take into account gravitational attraction between perturbations -- self-gravity remains radial, but since we limit ourselves to analyzing small perturbations at the initial stage, such attraction does not play any role.

After substitution (\ref{subsRho})-(\ref{subsVp}) to (\ref{parSysR})-(\ref{parSysP}) taking into account the unperturbed solution the variables may be split using $r^2\nabla^2 Y_l^m = -l(l+1)Y_l^m$ relation and we got the following equations for each mode (the indexes $l$ and $m$ are omitted): 
\begin{align}
&-\frac{\lambda-2}{\alpha} \Phi + x \Phi' + \frac{1}{x^2}\frac{d}{dx}x^2(\Phi f + \Upsilon g)- \nonumber \\
&  \;\;\;\;\;\;\;\;\;\;\; \;\;\;\;\;\;\;\;\;\;\; \;\;\;\;-\frac{l(l+1)}{x} g \Psi = 0, \label{forPhi} \\
& -\frac{\lambda+\alpha-1}{\alpha} \Upsilon + x \Upsilon' + \frac{d}{dx} \Upsilon f = \nonumber \\ 
 &\;\;\;\;\;\;\;\;\;\;\; \;\;\;\;\;\;\;\;\;\;\;\;\;\;\;\;\;\;\;\;\;\;= -\frac{1}{g}\Omega' + \frac{h'}{g^2}\Phi - \frac{1}{x^2}\Lambda, \label{forUps} \\
& -\frac{\lambda+\alpha-1}{\alpha}  \Psi + (f+x)\Psi' = -\frac{1}{x g}\Omega, \label{forPsi} \\
& \;\;\;\;\; \Lambda' = x^2 \Phi, \label{forM} \\
& -\frac{\lambda+2\alpha-4}{\alpha} \Omega + (f+x)\Omega' + h' \Upsilon + \nonumber \\
& \;\;\;\;\;\; + {\Gamma} h \left(\frac{1}{x^2}\frac{d}{dx}x^2 \Upsilon - \frac{l(l+1)}{x} \Psi\right) +{\Gamma} \Omega \frac{1}{x^2}\frac{d}{dx}x^2 f + \nonumber \\ 
& \;\;\;\;\;\;\;\;\;\;\; +({\Gamma}-1)g^{\kappa-1} h^{\chi-1}(\kappa h \Phi + \chi g \Omega) =0,  \label{forOmega}
\end{align}
where the prime denotes the $x$ derivative, $\kappa=s-n\sigma/\eta+1$, $\chi=n/\eta$. Now the partial differential equations (\ref{parSysR})-(\ref{parSysP}) are reduced to the system of the ordinary differential equations (\ref{forPhi})-(\ref{forOmega}) with the single self-similar variable $x=r/A (-t)^\alpha$ with fixed $\alpha,A>0$ (see (\ref{AspT})(\ref{ALPHAspT})). The derived equations do not contain $m$ therefore for each value of $l$ any $0\le m \le l$ is possible.

\subsection{\col{Asymptotic behaviour}}\label{sec:AsBeh}

\col{The self-similar variable $ x $ belongs to the interval $ 0 \le x < \infty $, which means that for any $ r \neq 0 $ the time varies in the interval $ - \infty <t \le 0 $, i.e.\@ the collapse begins at $ t = - \infty $ and ends at $ t = 0 $. Setting the boundary conditions corresponding to the absence of disturbances at $ x \rightarrow 0 $, we studied the behavior of the system (\ref{forPhi}) - (\ref{forOmega}) in this limit.}

\col{Since it was assumed that for $x\rightarrow 0$ the perturbations tend to zero and $g$ and $h$ reach a constant, in the equations (\ref {forPhi}) - (\ref {forOmega}) we left only terms containing derivatives of the perturbation amplitudes or terms proportional to the smallest negative degree of $ x $. As a result we got:
\begin{align}
&f \Phi' + g \Upsilon' + \frac{2}{x}( f\Phi + g\Upsilon ) - \frac{l(l+1)}{x}g \Psi = 0, \label{forPhi0} \\
&f\Upsilon' = -\frac{1}{g}\Omega' - \frac{1}{x^2}\Lambda,\label{forUps0} \\
&f\Psi' = -\frac{1}{x g} \Omega,\label{forPsi0} \\
&\Lambda' = 0, \label{forM0}\\
&f\Omega' + \colFo{\Gamma(0)} h \left( \Upsilon' + \frac{2}{x} \Upsilon - \frac{l(l+1)}{x}\Psi \right) + \frac{2}{x}\Gamma(0) f\Omega = 0. \label{forOmega0}
\end{align}}
\col{According to (\ref{forM0}) and the boundary conditions, we choose $\Lambda=0 $, which corresponds to the absence of self-gravity at $x=0$. \cite{Nadezhin1969} \colFo{showed} that $f\approx -2x/3 \alpha $ for $x\rightarrow 0 $, so we get $ \Omega=0 $ and only one equation \colFo{remains}:
\begin{align}
\colFo{	\Upsilon'+ \frac{2}{x}\Upsilon - \frac{l(l+1)}{x}\Psi = 0,}
\label{nearZeroUps}
\end{align}
\colFo{which indicates} the power-law behavior of $\Upsilon$ and $\Psi$ for $x\rightarrow 0$. Note that even without neglecting $f$ (but for $\Lambda = 0$) it is possible to search for a solution (\ref{forPhi0}) - (\ref{forOmega0}) in the form of a power-law ansatz $\sim x^k$, since all unknown functions enter into the equations either as derivatives or with factors $\sim x^{- 1}$. \colFo{Insofar as $x\sim r$, then in this limit the perturbations grow in a power-law manner, which is consistent with \cite{Lai2000}.}}

\colFo{
	Let us elaborate on the equations (\ref{forUps0})(\ref{forPsi0}) with $\Lambda=0$. Eliminating the pressure perturbations $\Omega$ and taking into account the above asymptotics, we arrive at the following expression:
	\begin{align}
		\colFo{	\Psi''+ \frac{2}{x}\Psi' - \frac{1}{x}\Upsilon' = 0.}
		\label{nearZeroPsi}
	\end{align}
Equation (\ref{nearZeroUps}) together with (\ref{nearZeroPsi}) give two possible solutions of velocity perturbations: $x^{l-1}$ and $x^{-l-2}$. }

\colFo{
	The fact that tangential perturbations of velocity $\Psi$ and radial one $\Upsilon$ grow near the center with the same law has an important physical meaning, apparently right for the entire flow as a whole. Indeed, let us consider $l\neq 0$ case because zero mode corresponds to $v_\theta=v_\phi=0$. Since the unperturbed flow is converging, a local increase or decrease in the radial flux in a certain solid angle leads to the appearance of azimuthal fluxes directed from (in the case of an increase) or to (a decrease) the place where the fluctuations arise. Due to this, a small perturbation can begin to develop, breaking the sphericity of the converging flow\colFo{, drawing energy from the gravitational field of the core}. Undoubtedly, this does not exhaust the description of the reasons for the development of the instability in such multi-component model. 
	Numerical modeling of CC in PISN progenitor can shed light on these reasons.}

\begin{figure*}[ht]
	
	\begin{minipage}{0.49\linewidth}
		\includegraphics[width=1\linewidth]{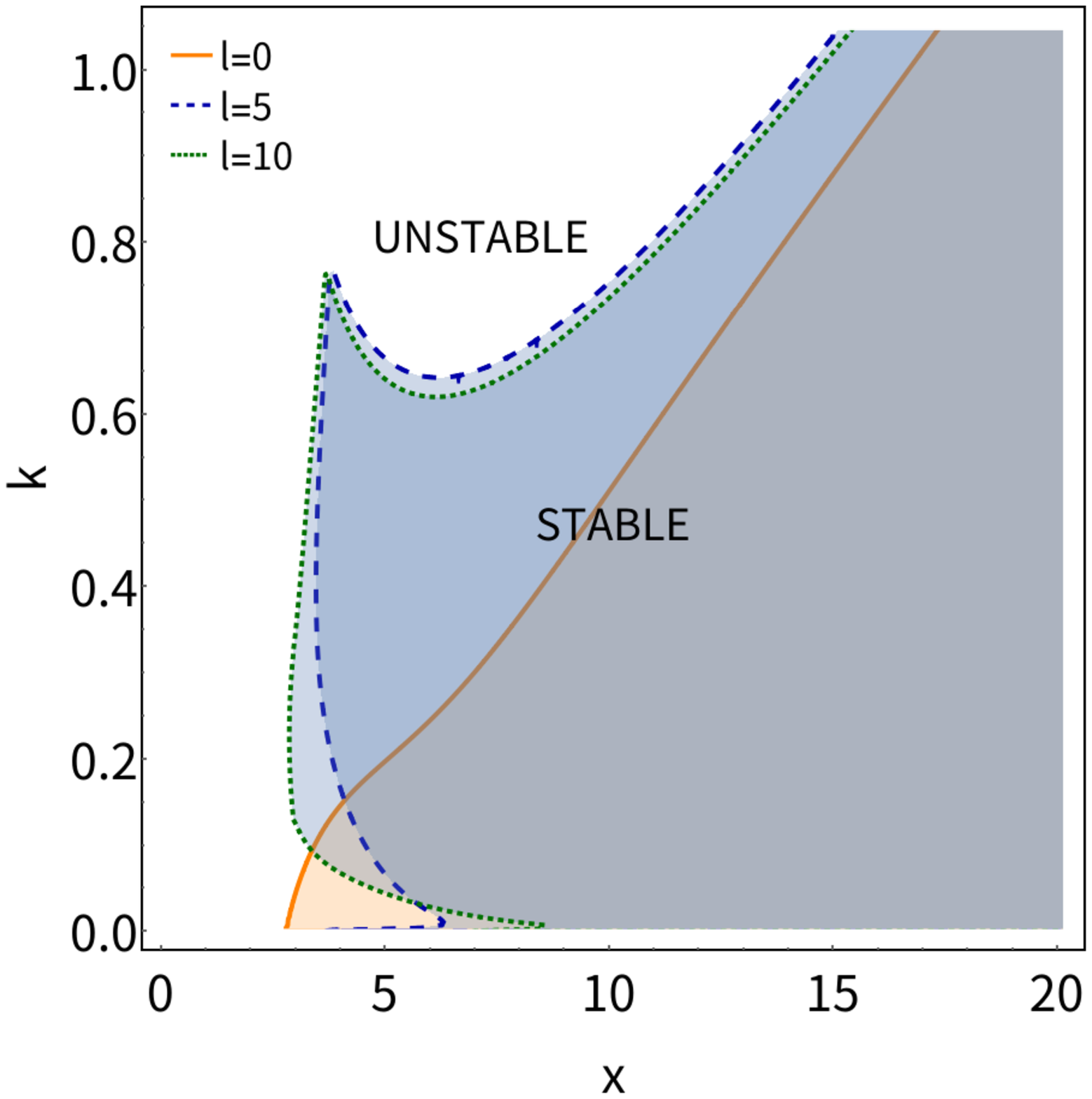}
	\end{minipage}
	\hfill
	\begin{minipage}{0.49\linewidth}
		\includegraphics[width=1\linewidth]{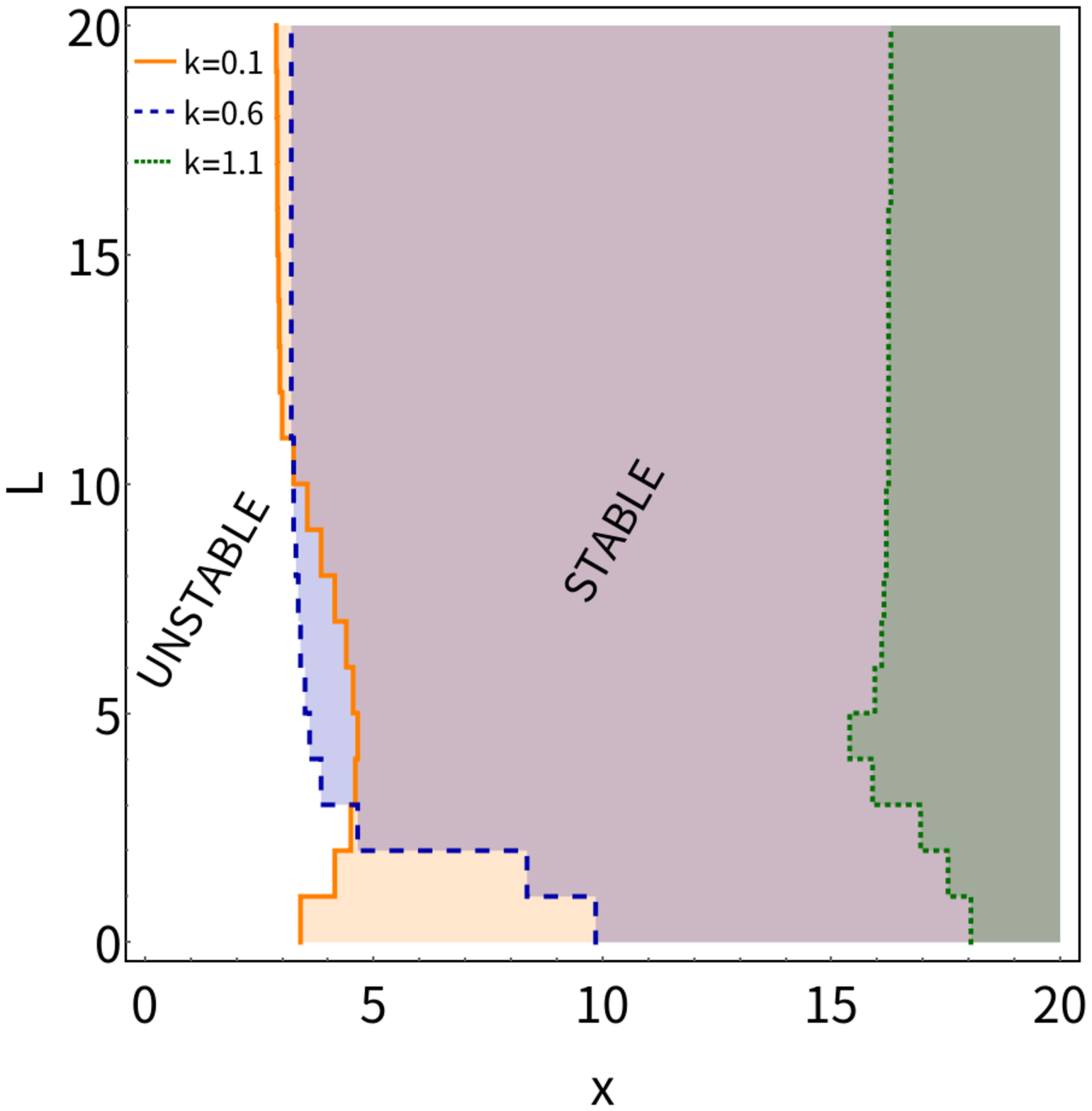}
	\end{minipage}
	\label{fig:stabB}
	\caption{Stability diagrams in $x-k$ (left panel) and $x-l$ (right panel) planes for model~A.}
	\label{fig:diagsA}
	
\end{figure*}

\smallskip
\subsection{Local stability} \label{subsec:local}

\begin{figure*}[ht]
	\begin{minipage}{0.49\linewidth}
		\includegraphics[width=1\linewidth]{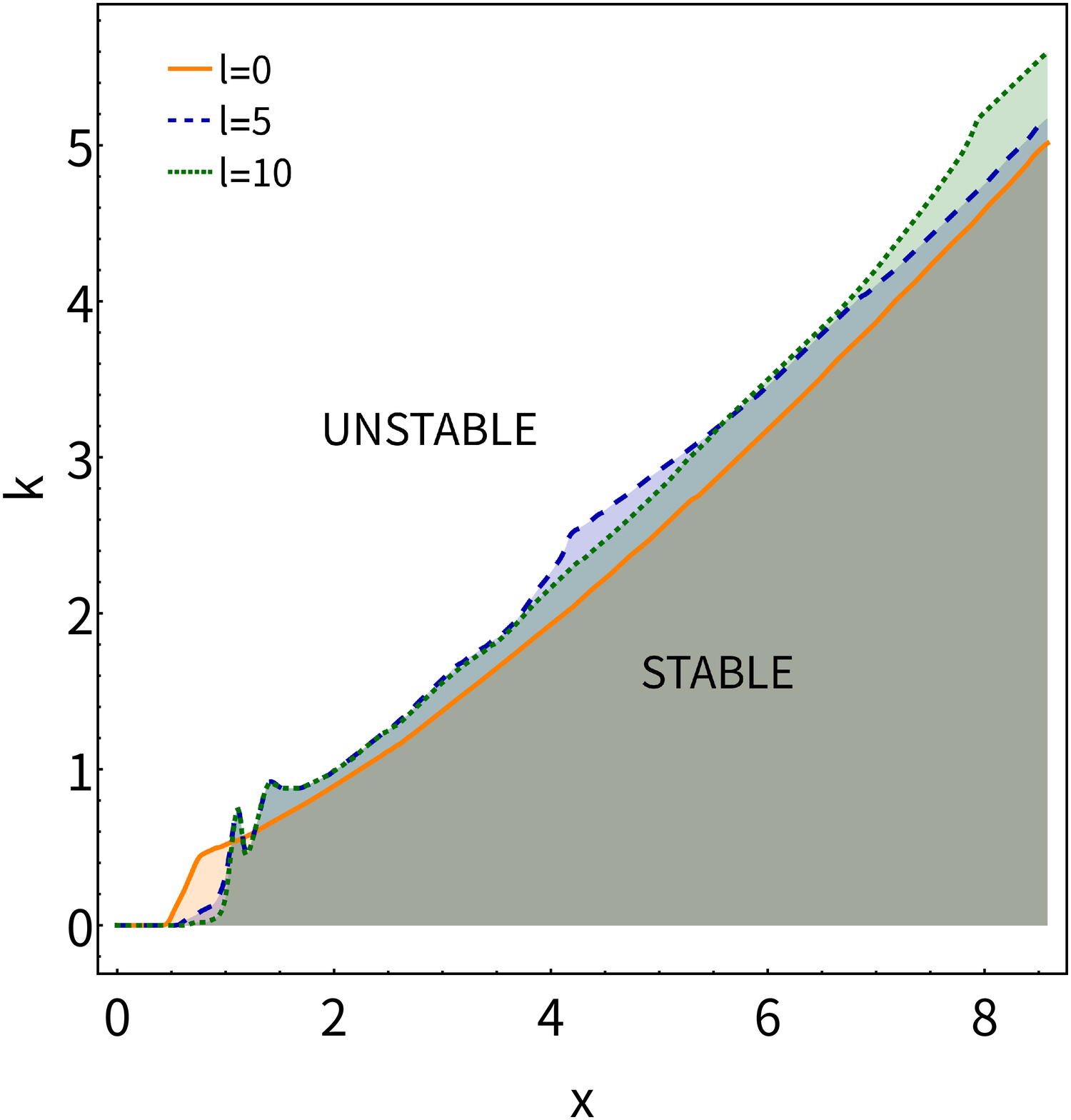}
	\end{minipage}
	\hfill
	\begin{minipage}{0.49\linewidth}
		\includegraphics[width=1\linewidth]{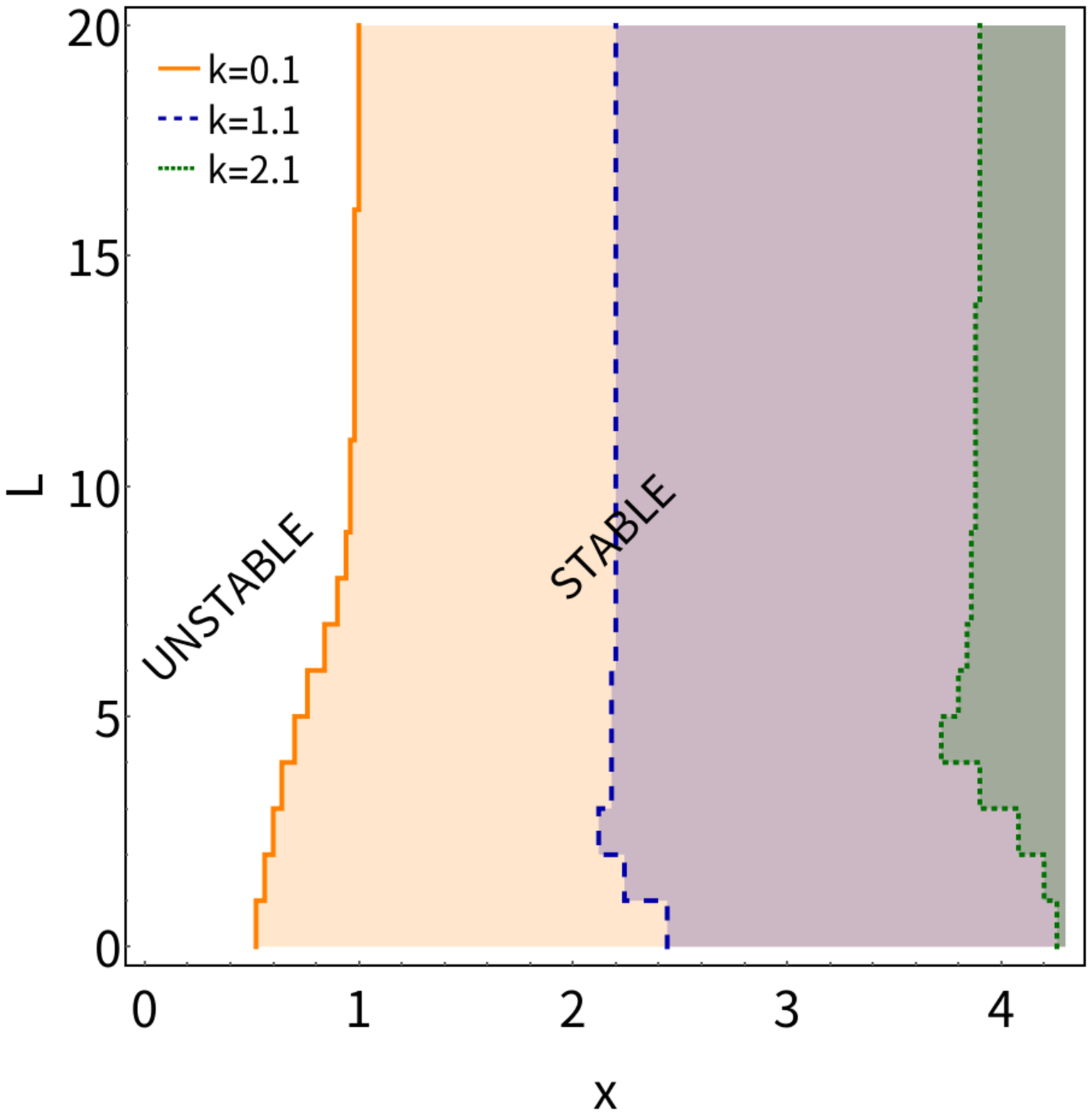}
	\end{minipage}
	\caption{Stability diagrams in $x-k$ (left panel) and $x-l$ (right panel) planes for model~B.}
	\label{fig:diagsB}
\end{figure*}

\col{In contrast to the works of \cite{Brushlinskii1982,murakami}, in which the stability of a converging spherical shock wave with boundary conditions corresponding to a strong discontinuity was studied, in our case taking into account the boundary conditions for perturbations seems insignificant, because losing stability occurs away from the boundaries (which is consistent with the results presented in Section \ref{subsec:diagsInf}). In such cases it is usually assumed (see e.g.\@ \cite{Ji2001}) that the perturbations propagate in the form of plane waves and from the analysis of the resulting dispersion equation, stability is concluded. In our case, due to the specificity of the obtained equations with the self-similar variable and their behavior near $x=0$, it was assumed that the perturbations change according to a power law. The passage to such a consideration is a strong assumption, but it is justified for considering stability in a small neighborhood of an arbitrary point $x$}. Such consideration may give us information about a point $x_*$, where the flow loses stability, however a functional behavior of the perturbations \col{on the entire interval $ 0 \le x <\infty $} stays unknown. 
Therefore, we represented the solution in the form: $\Phi(x) = \Phi_0 x^k$, $\Omega(x) = \Omega_0 x^k$, $\Lambda(x) = \Lambda_0 x^k$, $\Upsilon(x) = \Upsilon_0 x^k$, $\Psi(x) = \Psi_0 x^k$ and found $D_l(\lambda,k,x)$ -- a determinant of a system obtained after the substitution. It consists of all degrees of $\lambda$ up to forth degree, all degrees of $k$ up to fifth degree and complicated functions of $x$.

With respect to all our assumptions, the space-time behavior of the perturbations has the form: $(-t)^{\lambda-\alpha k} r^k$. Thus, we are interested in signature of the real part of $\mu=\lambda-\alpha k$, because it defines finite or infinite growing of the perturbations at $t\rightarrow 0$, therefore $\mu>0$ corresponds to the stable flow and $\mu<0$ to unstable. Also, we can restrict all possible values of $k$, because with a negative $k$ the perturbations near the center become larger than the main solution, so $k\ge 0$. The equation $D_l(\mu+\alpha k,k,x)=0$, solved relative of $\mu$, has four roots. We presumed that the flow is unstable if at least one of these roots has a negative real part. In this analysis we did not take into account the instabilities that may appear in the neighborhood of the singular point of the unperturbed system. They cannot have a physical meaning and their presence is connected with the inaccuracy of determining the conditions of passing the singular point.

\section{Analysis of results} \label{sec:Analysresult}

\subsection{Information from diagrams}\label{subsec:diagsInf}

\begin{figure*}[ht]
\begin{minipage}{0.49\linewidth}
\includegraphics[width=1\linewidth]{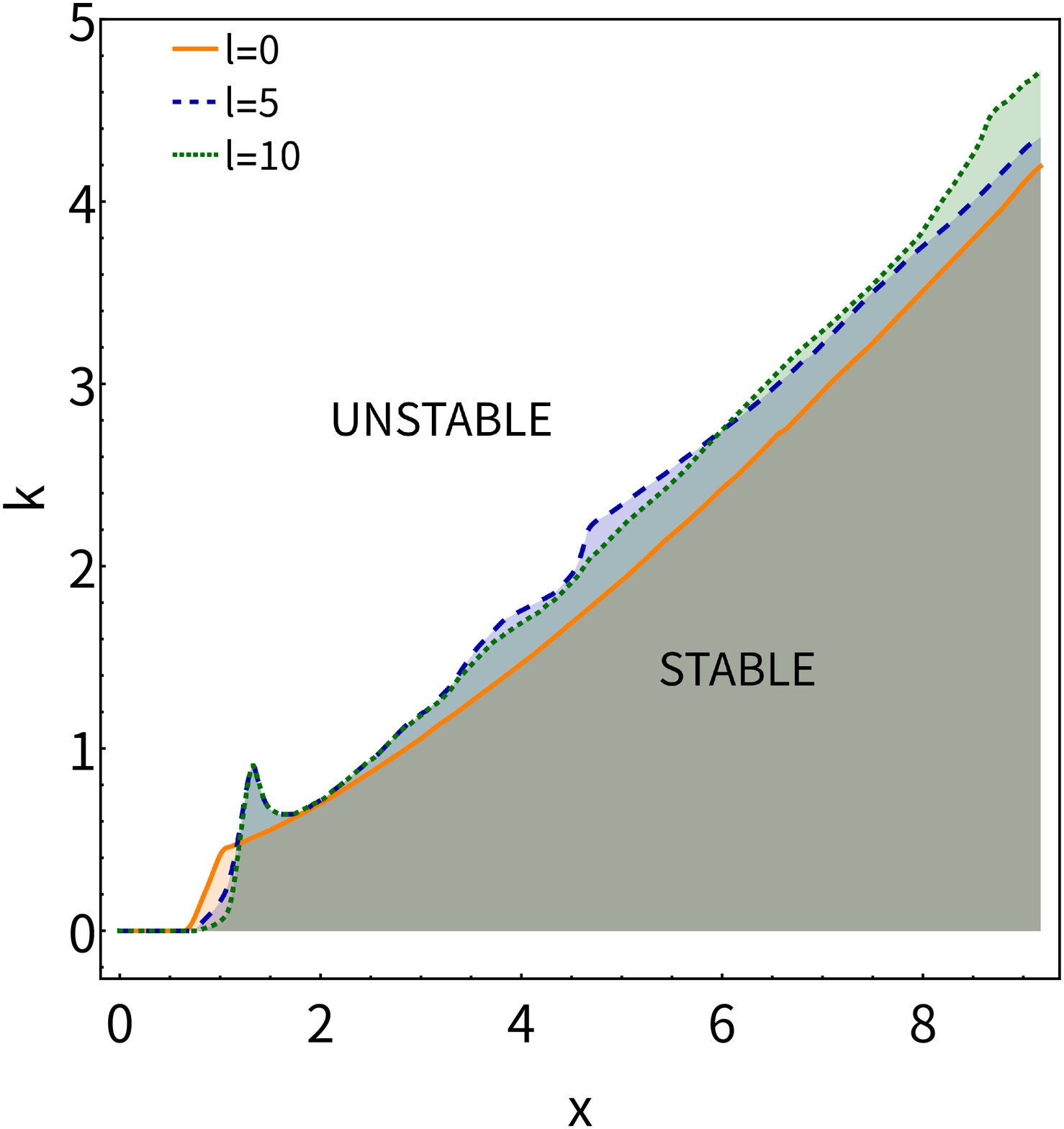}
\end{minipage}
\hfill
\begin{minipage}{0.49\linewidth}
\includegraphics[width=1\linewidth]{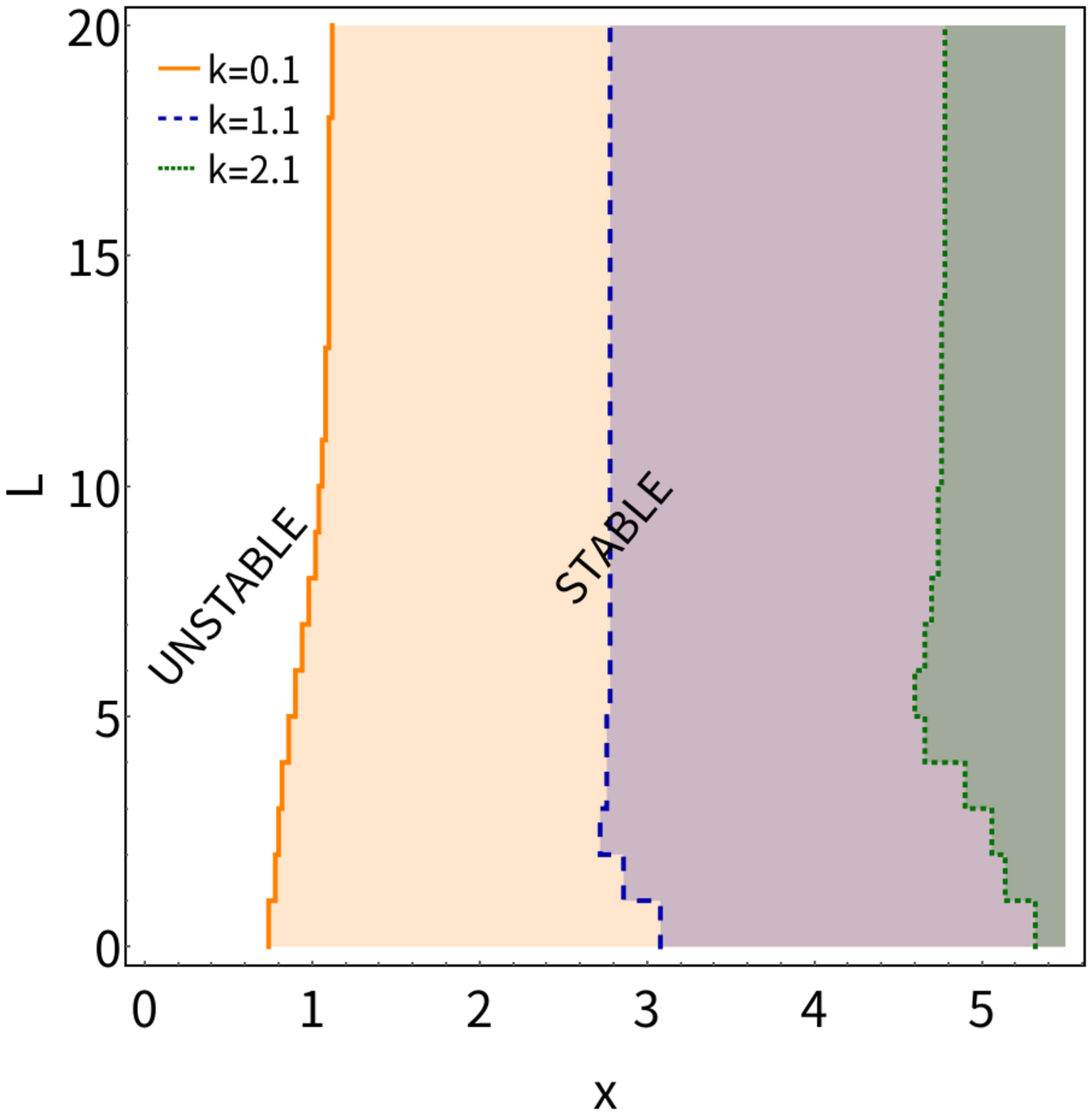}
\end{minipage}
\caption{Stability diagrams in $x-k$ (left panel) and $x-l$ (right panel) planes for model~C.}
\label{fig:diagsC}
\end{figure*}

\col{Solved equation $D_l(\mu+\alpha k,k,x)=0$ defines a two-dimensional surface in the parameter space $(x,k,l)$, separating stable and unstable zones. Different sections of the parametric space are shown in Figures~\ref{fig:diagsA},\ref{fig:diagsB},\ref{fig:diagsC} for models~A,~B,~C respectively. }

Firstly let us consider the stability diagrams for model~A. As may be seen from the diagram in $x-k$ plane in Figure~\ref{fig:diagsA}, the zeroth harmonic gets unstable at $x_*\approx 2.9$ with minimum value of $k$. Next, the value of $x_*$ grows almost linearly with growing $k$. For the fifth mode, the picture is different -- the stable zone has non-monotonic shape at small values of $x$, then it expands linearly. The stable area of $l=10$ mode demonstrates almost the same behavior but \col{it is slightly shifted to the low values of $x$}. 

\col{From the diagram in $x-l$ plane in} Figure~\ref{fig:diagsA} we can see that for each $k$ the shapes of the areas have the key feature -- almost vertical line at position $x=x_w$, value of $x_*$ leads to it with growing $l$. It means that at $x=x_w$ all harmonics, which are large or equal than a certain $l_w$, lose stability. For each value of $k$ the position of $x_w$ is different. For sufficiently large $k$ the instability begins with large values of $l$. Therefore, we can assume that at a point $x=x_w$ a harmonic with $l=l_w$ loses stability, harmonics with $l<l_w$ are stable at this point and harmonics with $l>l_w$ lose stability too. \colFo{But the loss of stability by higher than $l_w$ modes is insignificant} because later on, the system will transfer to nonlinear stage and the lower-energy and lower-scale perturbations $l>l_w$ will be absorbed by the higher perturbation $l=l_w$. To estimate $x_w$ we compared energy of these perturbations, at which point they lose stability. $l$ mode loses stability in a point $x_l$ having unit kinetic energy $e_l \sim {\rho_\text{in}} \dot{R}^2 x_l^{3k} $. We supposed that $x_w$ is achieved if the difference of the energy between $l$ and $l+1$ modes is smaller than the characteristic energy ${\rho_\text{in}} \dot{R}^2$, i.e.\@ $|x_{l+1}^{3k} - x_l^{3k}|\ll 1$. According to Figure~\ref{fig:diagsA} for $k=0.1$ this relation takes place for zeroth mode at $x_w\approx 3.5$; for $k=0.5$ it is $l_w\approx 8$ at $x_w\approx 3.5$ and for $k=1$ we approximately have $l\approx 15$  at $x_w\approx 14.5$.

\col{As it may be seen from Figures~\ref{fig:diagsB},\ref{fig:diagsC} the stability diagrams for models~B~and~C are almost the same, although the functional behavior of $\Gamma$ for these models is quite different. In Figure~\ref{fig:diagsB} we may conclude that for model~B the zeroth harmonic is stable for low $k$ until $x_*\approx 1$. For higher modes the most stable point shifts to $x_*\approx 2$. Analyzing the diagrams in $x-l$ plane and making the energy estimations by the same way for each calculated $k$ we have $x_w\approx 1$, $l_w\approx 5$ for $k=0.1$; $x_w\approx 2.1$, $l_w\approx 3$ for $k=1.1$ and $x_w\approx 3.9$, $l_w\approx 4$ for $k=2.1$. Similar estimations for model~C according to Figure~\ref{fig:diagsC} give $x_w\approx 1$, $l_w\approx 8$ for $k=0.1$; $x_w\approx 2.8$, $l_w\approx 2$ for $k=1.1$ and $x_w\approx 4.6$, $l_w\approx 4$ for $k=2.1$.}

It can be seen that there are no conditions under which the flow would remain stable up to the moment of collapse, \colFo{which also follows from the asymptotic analysis in Section~\ref{sec:AsBeh}}. Since the functional behavior remains unknown under our consideration, then to make some estimations we have to choose an appropriate value of $k$. For each of the models considered, we chose such $k$, which corresponds to the most stable regime. As can be seen from Figure~\ref{fig:diagsA}, for model~A it is about $k\approx 0.25$ with $x=x_*\approx 3.5$. \col{For models~B~and~C, according to Figures~\ref{fig:diagsB},\ref{fig:diagsC}, it is $k\approx 0.5$ with $x_*\approx 1$.}

\subsection{Physical consequences} \label{subsec:phyRes}

At the distance $r$ losing stability occurs at ${\tau_\text{inst}}=(r/x_* A)^{1/\alpha}$ before the explosion. To find out the possibility of the development of the instabilities inside a real star, let us compare the time ${\tau_\text{inst}}$ with the time of free fall ${\tau_\text{ff}}=(G\rho)^{-2}$. Despite the fact that the dominant neutrino emission \colFo{and transport} processes are different in \colFo{massive and} very massive stars, for completeness we made the corresponding estimates for both types of stars.

For the considered parameters for model~A, we have $\alpha=0.9$ and the self-similar amplitude is $A\sim 10^9 \text{ cm s}^{-0.9}$ (\cite{Murzina1991}). With the conditions inside the core of a very massive ($\sim 200M_\odot$) star $\rho_c\sim 10^5 \text{ g }\text{cm}^{-3}$, $r_c\sim 0.1 R_\odot$, we have ${\tau_\text{ff}}\sim 10 \text{ s}$ and ${\tau_\text{inst}} \sim 1 \text{ s}$ which means that the perturbations have enough time to develop. Now let us make the same estimation for a massive star. Typical parameters for the core of a $10 M_\odot$ star are: $\rho_c\sim 10^9 \text{ g} \text{ cm}^{-3}$, $r_c \sim 10^{-3} R_\odot$. Thus, we get ${\tau_\text{ff}} \sim {\tau_\text{inst}} \sim 10^{-2} \text{ s}$, which does not guarantee the availability of sufficient time for the development of the instability.

\col{\colFo{The core of PISN progenitor at the latest stage mostly consists of oxygen or other elements of a similar mass (\cite{Rakavy1967}), therefore} for our estimations it is enough to calculate $A$ using (\ref{AspT}) supposing only oxygen stellar matter. Thus for models~B~and~C, we have $A\simeq 10^9\text{ cm s}^{-0.75}$. \colFo{For very massive stars it} again gives ${\tau_\text{inst}} \sim 1 \text{ s} \ll \tau_\text{ff}$, which means the possibility of the development of instability  during the collapse of the core of a massive star. }

Knowing an angular mode $l_w$ of the perturbation we are able to judge about the numbers of initial ignition spots, assuming that they appear at locations of extremes of a spherical harmonic. A spherical harmonic $Y^m_l$ has $l$ local extremes, which are located symmetrically around the equator $\theta=\pi/2$ in case of $m=0$. Since the index $m$ is not included in equations (\ref{forPhi})-(\ref{forOmega}) all values of $m\leq l$ are possible for each fixed $l$. Therefore at the moment when the flow gets unstable, there are $l_w$ points where the instability develops, located as extremes of $Y^m_l(\theta,\phi)$. Of course later, at nonlinear stages, this amount may change because of any kind of interaction of different part of a core and at the moment of explosion the numbers of ignition spots may be different, but, as we believe, it must be on order of $l_w$. \col{In case of models~B~and~C, which correspond to \colFo{CC} of PISNe progenitor, we have $l_w\approx 5$.} 

\colFo{\cite{Mller2016} demonstrated with numerical simulations of $18M_\odot$ CC the appearance for some time of $l=3$ and $l=4$ modes in the oxygen envelope. Modeling of CC in more massive stars ($85M_\odot$, $100M_\odot$) done by \cite{Powell2021} reveals the formation of similar structures. Thus, despite the fact that these results refer to less massive stars than PISN progenitors, and therefore have other physical processes (e.g.\@ neutrino transport), they also demonstrate the onset of instability during collapse.}

\colFo{Knowledge of the disturbances mode} allowed us to estimate the size of the perturbations, which may be used in numerical simulations as a characteristic size of the spots. If the loss of stability occurs at $r\sim r_c$ and there are $\approx 5$ ignition spots at this position then each spot takes $2\pi r_c /5$, i.e.\@ radius of a spot must be $\sim \pi r_c /5$.

Let us check that the got results are in the limits of applicability of the model. In the work of \cite{Murzina1991}, these limits were established, they are due to the appearance of the opacity of the collapsing matter for the radiation that carries energy away. In this work it is concluded their model (model~A in our definitions) is applicable for $x=3.5$ if $|t/0.025 \text{ s}|>0.24$, what is performed for $t={\tau_\text{inst}}$. For \col{models~B~and~C with $x=1$ the same calculations give $|t/4.92\cdot 10^{-9} \text{ s}|>0.68$ and $|t/2.63\cdot 10^{-9} \text{ s}|>0.42$ respectively.} Thus, it turns out that our consideration \cols{does not go beyond the scope of the approximation}.

\colFo{The initial model does not include differential rotation of PISN progenitor, which overlooks the possible occurrence of instability associated with the differential rotation of the stratified fluid. Also, taking into account the rotation will lead to the dependence of the perturbations on $m$ number as result of Coriolis and centrifugal forces. A more accurate account of EoS should lead to a decrease in pressure and, as a consequence, to a slight suppression of instability due to equations (\ref{forUps0})(\ref{forPsi0}). Herewith, the correct allowance for neutrino cooling should promote the growth of instabilities, since the steeper behavior of the function $Q_\nu$ will lead to a larger deviation of small perturbations. However, it does not seem possible to include the effects presented above in the self-similar consideration.}

\section{Summary} \label{sec:concl}

We presented results of our local analysis of \colFo{CC} stability in frames of the self-similar model.
\col{By slightly modifying the model of \cite{Nadezhin1969}, we included in our consideration the possibility of changing the adiabatic exponent $\gamma$ in space and time \cols{as a self-similar function} to get the model describing PISNe  progenitor. In addition to the already known solution describing the neutrino emission due to Urca process (model~A), having set two definite laws of $\gamma$ changing (Figure~\ref{fig:gammas}) and the law of neutrino emission according to \cite{schinder}, we found two new solutions (models~B~and~C, see Table~\ref{tab:params} \cols{and Figure~\ref{fig:SolABC}}) that are quite consistent with the results of \cols{one dimensional} numerical calculations \colFo{both done by us and} presented in the work of \cite{Chatzopoulos2013} (see Figure~\ref{fig:physBC}). 
	\cols{The form of $\gamma$ for model~C was chosen to match the numerical results given \colFo{by} \cite{Gilmer2017} as best as possible, while the behavior of $\gamma$ for model~B is illustrative.}
	Investigating the obtained solutions for stability with respect to three-dimensional small perturbations, we got the stability diagrams for each of the solutions (Figures~\ref{fig:diagsA},\ref{fig:diagsB},\ref{fig:diagsC}) using those we can judge about conditions of collapse stability. 
} \cols{Comparing the diagrams for models~B~and~C, we can conclude that the stability of the collapse in the model under consideration is \colIK{almost insensitive} to reasonable changes~of~$\gamma$.}

Analyzing the results, we may conclude there are no conditions under which the collapse inside \colFo{very} massive ($\sim 200M_\odot$) stars might be stable, while the \colFo{CC} of \colFo{massive} ($\sim 10M_\odot$) stars must be stable.
Estimating the energy of modes, we concluded that the most danger mode for models~B~and~C is mode with $l=5$. It means that when the flow loses stability, there are $\simeq 5$ areas where the initial perturbations may continue to develop. 

\cols{On the other hand, as it is shown \colFo{by} \cite{Baranov2013}, introducing thermal energy into the central region of a star in the form of a series of multiple hot spots \colIK{as an initial condition} leads to a multi-core explosion. \cite{Chardonnet2015} showed that nonuniform explosions may be responsible for some of the observed GRBs. We assume that such hot spots can be formed as a result of found instability at the collapse stage. Although full three-dimensional numerical modeling of \colFo{CC} is necessary to prove this assumption, we hope that the results obtained from the simple analytical model indicate what to expect from it.}

\col{The fact that the stability analysis for models~B~and~C, which have enough different $\gamma$ behavior, got almost identical results \cols{of stability investigation}, raises a question. Probably, very massive stars which will not undergo PISNe \cols{scenario}, \colFo{may} also develop such instability, because high temperature implies neutrino losses. Normally, such stars results in black holes in the spherical scenario. Will such instability change their fate? We think that this question requires a separate study.}

\begin{acknowledgements}
I. Kalashnikov would like to thank Theoretical Physics and Mathematics Advancement Foundation ''BASIS'' for the financial support and LAPTh for their kind hospitality.
P. Chardonnet acknowledges the support of MEPhI where the work was initiated.
\end{acknowledgements}

\bibliography{biblio}

\end{document}